\documentclass[twocolumn]{aastex631}

\usepackage{epsfig}
\usepackage{graphicx}
\usepackage{amsthm}
\usepackage{amssymb}
\usepackage{rotating}
\usepackage{multirow}
\usepackage{longtable}
\usepackage{savesym}
\savesymbol{tablenum}
\restoresymbol{SIX}{tablenum}
\usepackage{booktabs}
\usepackage{mathtools}
\newcommand{\RNum}[1]{\uppercase\expandafter{\romannumeral #1\relax}}

\defcitealias{2019ApJ...873...51L}{Paper~I}
\defcitealias{2020ApJ...888...98L}{Paper~II}
\defcitealias{2021ApJ...923..198D}{Paper~III}

\accepted{\today}

\submitjournal{ApJ}

\shorttitle{Surveying G\ion{H}{2} Regions: IV. Sgr\,D, W42, and a Reassessment of the Census}
\shortauthors{De Buizer}

\begin{document}

\title{Surveying the Giant \ion{H}{2} Regions of the Milky Way with SOFIA: IV. Sgr\,D, W42, and a Reassessment of the Giant \ion{H}{2} Region Census}

\correspondingauthor{James De Buizer}
\email{jdebuizer@sofia.usra.edu}

\author[0000-0001-7378-4430]{James M. De Buizer}
\affil{SOFIA-USRA, NASA Ames Research Center, MS 232-12, Moffett Field, CA 94035, USA}

\author[0000-0003-4243-6809]{Wanggi Lim}
\affil{SOFIA-USRA, NASA Ames Research Center, MS 232-12, Moffett Field, CA 94035, USA}

\author[0000-0003-3682-854X]{Nicole Karnath}
\affil{SOFIA-USRA, NASA Ames Research Center, MS 232-12, Moffett Field, CA 94035, USA}

\author[0000-0003-0740-2259]{James T. Radomski}
\affil{SOFIA-USRA, NASA Ames Research Center, MS 232-12, Moffett Field, CA 94035, USA}

\author[0000-0002-0915-4853]{Lars Bonne}
\affil{SOFIA-USRA, NASA Ames Research Center, MS 232-12, Moffett Field, CA 94035, USA}

\begin{abstract}
This is the fourth paper exploring the infrared properties of giant \ion{H}{2} regions with the FORCAST instrument on the Stratospheric Observatory For Infrared Astronomy (SOFIA). Our survey utilizes the census of 56 Milky Way giant \ion{H}{2} regions identified by \citet{2004MNRAS.355..899C}, and in this paper we present the 20 and 37\,$\mu$m imaging data we have obtained from SOFIA for sources Sgr\,D and W42. Based upon the SOFIA data and other multi-wavelength data, we derive and discuss the detailed physical properties of the individual compact sources and sub-regions as well as the large scale properties of Sgr\,D and W42. However, improved measurements have revealed much closer distances to both regions than previously believed, and consequently both sources are not powerful enough to be considered giant \ion{H}{2} regions any longer. Motivated by this, we revisit the census of giant \ion{H}{2} regions, performing a search through the last two decades of literature to update each source with the most recent and/or most accurate distance measurements. Based on these new distance estimates, we determine that 14 sources in total (25\%) are at sufficiently reliable and closer distances that they are not powerful enough to be considered giant \ion{H}{2} regions. We briefly discuss the observational and physical characteristics specific to Sgr\,D and W42 and show that they have properties distinct from the giant \ion{H}{2} regions previously studied as a part of this survey.
\end{abstract}


\section{Introduction} 

This is the fourth paper in a series of studies of the properties of Milky Way  giant \ion{H}{2} (G\ion{H}{2}) regions which  represent the largest and most powerful star forming areas of the Galaxy. Regions like these dominate the emission of most galaxies, and therefore G\ion{H}{2} regions are laboratories for understanding large-scale star-formation within galaxies in general \citep{1990ARA&A..28..525S}.  As their moniker implies, the large clusters of high-mass stars and protostars within G\ion{H}{2} regions provide a tremendous supply of ionizing (i.e., Lyman continuum) photons which create vast, bright, and optically thin cm radio continuum regions. For example, M17 has a radio continuum diameter of over 10\,pc and contains $\sim$100 O and B-type stars \citep{2008ApJ...686..310H}. 

Because these regions are so bright in the mid-infrared, the highest spatial resolution images from space missions like Spitzer Space Telescope and the Wide-field Infrared Survey Explorer (WISE) are saturated over most of their emitting areas at wavelengths $\ge$15\,$\mu$m. While some sources and sub-regions within some G\ion{H}{2} regions have been imaged at (sub-)arcsecond resolution via ground-based mid-infrared facilities \citep[e.g.,][]{2000ApJ...540..316S,2016ApJ...825...54B,2002AJ....124.1636K}, there are technical limitations (like small fields of view) that make it impractical to make the large maps required to fully cover the entire mid-infrared emitting area of these G\ion{H}{2} regions, many of which are more than 4$\arcmin$ in diameter. To date, the most complete maps in the mid-infrared are from the Midcourse Space Experiment (MSX), which have a spatial resolution of only $\sim$18$\arcsec$ at 22\,$\mu$m. Our survey source list comes from the census of \citet{2004MNRAS.355..899C} who identified 56 G\ion{H}{2} regions in our Galaxy based on their 6\,cm radio and mid-infrared (via MSX data) fluxes. Though this census contains the brightest star-forming regions in the Galaxy, it is not considered a complete survey of the entire population of radio G\ion{H}{2} regions. Nevertheless, the ultimate goal of our project is to compile a 20 and 37\,$\mu$m imaging survey of as many G\ion{H}{2} regions within the Milky Way as we can with the Stratospheric Observatory For Infrared Astronomy (SOFIA) and its mid-infrared instrument FORCAST, creating complete and unsaturated maps of these regions with the best resolution ever achieved at our longest wavelength (i.e., $\sim$3$\arcsec$ at 37\,$\mu$m). From our infrared observations we will gain a better understanding of their physical properties individually and as a population. 

In this paper we will, in part, concentrate on the SOFIA data we have obtained for sources Sgr\,D and W42. Though these sources are from the G\ion{H}{2} region census from \citet{2004MNRAS.355..899C}, more recent and/or more accurate measurements of the distances to both of these sources have shown that both are much closer to the Sun than previously believed. G\ion{H}{2} regions possess a Lyman continuum photon rate, $N_{LyC}$, of greater than $10^{50}$ photons/s \citep{2004MNRAS.355..899C, 1970IAUS...38..107M}, and the derivation of this value is dependent upon source distance. We will show in this paper that recalculating $N_{LyC}$ using the new, closer distances for both Sgr\,D and W42 indicate that neither qualifies as a G\ion{H}{2} region under the above criterion. 

As these regions are not G\ion{H}{2} regions, we will not go into the same depth of analyses as we have for our previous papers in this survey. However, since we have obtained SOFIA data on both of these regions, we were motivated to see what could be learned by comparing and contrasting their properties to those of the most luminous G\ion{H}{2} regions we have previously studied. Therefore the first goal of this paper is to derive the physical properties for the individual sources and sub-regions that make up both Sgr\,D and W42, and to quantify the nature of these regions as a whole.  

However, given the importance of distance on the classification of our source list as bona fide G\ion{H}{2} regions, and given the almost two decades that have past since that census was published, we were compelled to perform a literature search for the remaining \citet{2004MNRAS.355..899C} sources to compile the latest distance estimates for each source and recalculate their Lyman continuum photon rates to see which sources still qualify as G\ion{H}{2} regions. The distances to most of these sources had previously only been determined through kinematic methods, which are notoriously unreliable. Being sites for the formation of some of the most massive stars in the Galaxy means that G\ion{H}{2} regions often contain massive young stellar objects (MYSOs) displaying maser activity, and this has led to a host of studies in the past decade that utilize the masers to obtain accurate trigonometric parallax measurements to such regions. Other methods, like spectrophotometry as well as stellar parallax measurements from the Global Astrometric Interferometer for Astrophysics mission (GAIA), have also been utilized to obtain more accurate distance measurements to these regions in the almost two decades since \citet{2004MNRAS.355..899C}. Therefore, the second goal of this paper is to create an updated census of G\ion{H}{2} regions to serve as the source list for our survey moving forward. 

This paper is organized in the following manner. In Section~\ref{sec:data}, we will discuss the SOFIA observations and data reduction and analyses for both Sgr\,D and W42, including the creation and modeling of the infrared SEDs for sources within each region. We will give more background on both regions as we compare our new data to previous observations and briefly discuss the nature of each region and discuss individual sources and sub-regions in-depth in Sections~\ref{sec:SgrD} (Sgr\,D)  and \ref{sec:W42} (W42). In Section~\ref{sec:reassess} we discuss in detail the reassessment of the \citet{2004MNRAS.355..899C} census, including discussions of the methods used to determine distances to the entire list of G\ion{H}{2} regions, explanation of the calculation of Lyman continuum photon rate, and discussion of details related to the final list of updated G\ion{H}{2} regions. In Section~\ref{sec:reject}, we examine the nature of rejected G\ion{H}{2} region sources \object{Sgr D} and \object{W42} and investigate the their physical properties (beyond just Lyman continuum photon rate) and how they compare to the sources we have previously studied with SOFIA that are bonafide G\ion{H}{2} regions. We summarize our results in Section~\ref{sec:sum}.   

\section{Observations, Data Reduction, and Analyses}\label{sec:data}
 The SOFIA data for these sources were obtained in the same manner as for those in our previous three papers, and we direct the reader to the discussion of those details in \citetalias{2019ApJ...873...51L} (i.e., Lim et al. 2019). We will highlight below some of observation and reduction details specific to the Sgr\,D and W42 observations. 

The data for Sgr\,D were obtained during SOFIA's Cycle 3 using the FORCAST instrument \citep{2013PASP..125.1393H} on the night of 2015 June 24 (Flight 221). FORCAST is a dual-array mid-infrared camera capable of taking simultaneous images at two wavelengths. The short wavelength camera (SWC) is a 256$\times$256 pixel Si:As array optimized for 5–-25\,$\mu$m observations; the long wavelength camera (LWC) is a 256$\times$256 pixel Si:Sb array optimized for 25–-40\,$\mu$m observations. After correction for focal plane distortion, FORCAST effectively samples at 0$\farcs$768 pixel$^{-1}$, which yields a 3$\farcm$4$\times$3$\farcm$2 instantaneous field of view. Observations of Sgr\,D were obtained using the 20\,$\mu$m ($\lambda_{eff}$=19.7\,$\mu$m; $\Delta\lambda$=5.5\,$\mu$m) and 37\,$\mu$m ($\lambda_{eff}$=37.1\,$\mu$m; $\Delta\lambda$=3.3\,$\mu$m) filters simultaneously using an internal dichroic. Data were obtained at aircraft altitude of 36,000 feet by employing the standard chop-nod observing technique used in the thermal infrared, with 3$\arcmin$ chop and nod throws and an on-source integration time of about 450s in both filters. 

For W42, data were obtained during SOFIA's Cycle 2 on the night of 2014 June 4 (Flight 176). A different filter was used in the short wavelength camera of FORCAST for these earlier observations of our project, namely the 25\,$\mu$m ($\lambda_{eff}$=25.3\,$\mu$m; $\Delta\lambda$=1.9\,$\mu$m) filter. However, the dichroic mode was still employed so that these observations were taken simultaneously with the same 37\,$\mu$m filter as was used for the Sgr\,D observations. Data were obtained at aircraft altitude of 43,000 feet with 5$\arcmin$ chop and nod throws. Unlike Sgr\,D, the mid-infrared emitting region of W42 is larger ($\sim$4.5$\arcmin\times$4.5$\arcmin$) than the FORCAST field of view, and thus had to be mapped using multiple pointings. We created a mosaic from 3 individual pointings, with two pointings having an average on-source exposure time of about 160s at both 25\,$\mu$m and 37\,$\mu$m, and the southernmost pointing only having an on-source exposure time of 35s (due to telescope issues cutting the observation short). Images from each individual pointing were stitched together into a final mosaic using the SOFIA Data Pipeline software REDUX \citep{2015ASPC..495..355C}. 

Flux calibration for all observations were provided by the SOFIA Data Cycle System (DCS) pipeline and the final total photometric errors in the mosaic were derived using the same process described in \citetalias{2019ApJ...873...51L}. The estimated total photometric errors are 15\% for 20\,$\mu$m and 10\% for 37\,$\mu$m. All images then had their astrometry absolutely calibrated using Spitzer-IRAC data by matching up the centroids of point sources in common between the Spitzer and SOFIA data. Absolute astrometry of the final SOFIA images is assumed to be better than 1$\farcs$5. The effective spatial resolution of the data is $\sim$2.5$\arcsec$ at 20 and 25\,$\mu$m and $\sim$3.2$\arcsec$ at 37\,$\mu$m. 

\begin{figure}[tb!]
\begin{center}
\includegraphics[width=3.3in]{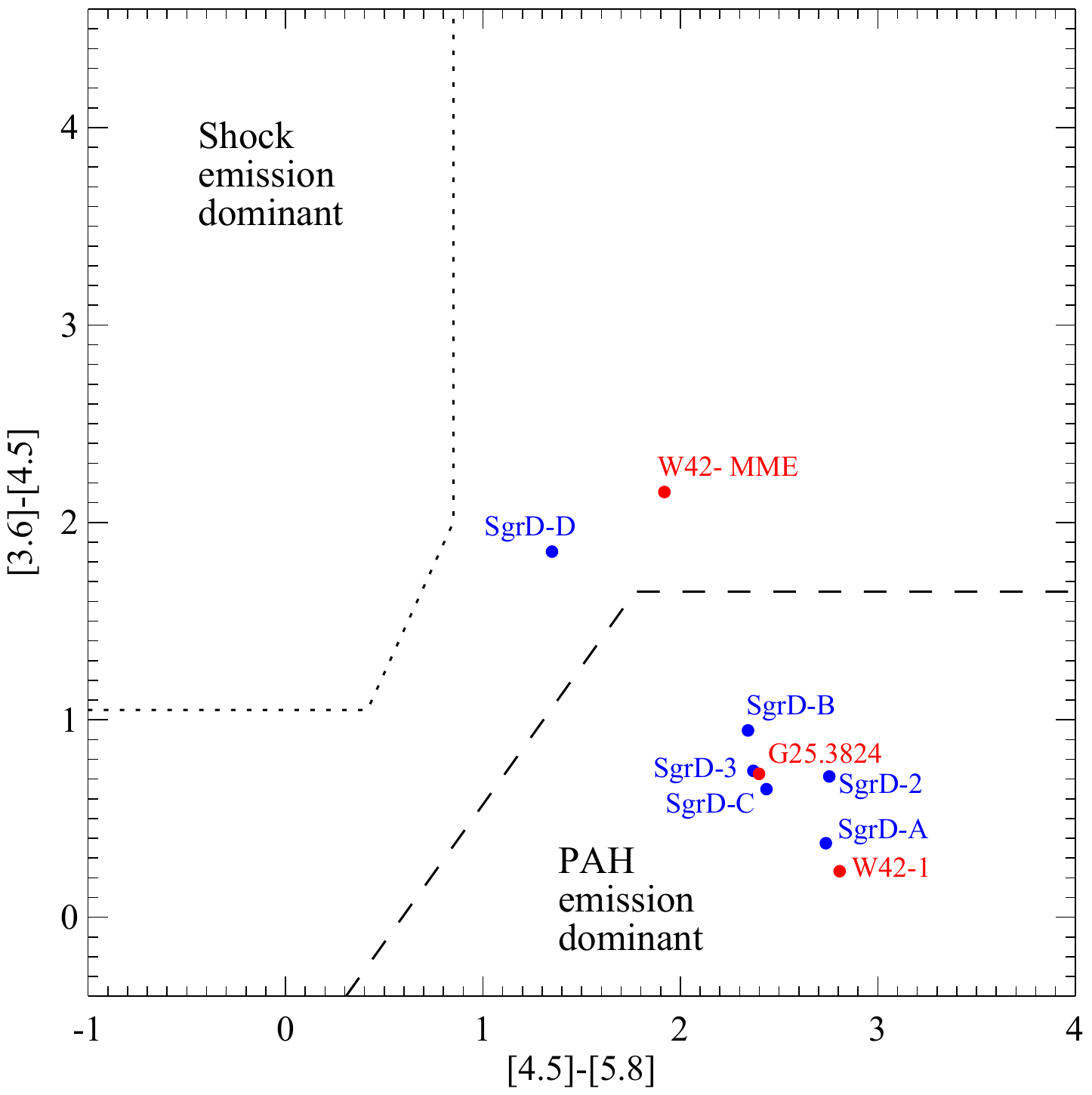}\\
\end{center}
\caption{\footnotesize A color-color diagram utilizing the background-subtracted Spitzer-IRAC 3.6, 4.5, and 5.8\,$\mu$m source photometry to distinguish ``shocked emission dominant'' and ``PAH emission dominant'' YSO candidates for the compact sources within Sgr\,D (blue) and W42 (red). Above (up-left) of dotted line indicates shock emission dominant regime. Below (bottom-right) dashed line indicates PAH dominant regime. We adopt this metric from \citet{2009ApJS..184...18G}.}
\label{fig:cc}
\end{figure}

\begin{deluxetable*}{rrrrrrrrrrrrr}
\centering
\tabletypesize{\scriptsize}
\tablecolumns{8}
\tablewidth{0pt}
\tablecaption{SOFIA Observational Parameters for Sources in Sgr\,D and W42}\label{tb:obspar1}
\tablehead{\colhead{  }&
           \colhead{  }&
           \colhead{  }&
           \multicolumn{3}{c}{${\rm 20\mu{m}}$}&
           \multicolumn{3}{c}{${\rm 25\mu{m}}$}&
           \multicolumn{3}{c}{${\rm 37\mu{m}}$}&\\
           \cmidrule(lr){4-6} \cmidrule(lr){7-9} \cmidrule(lr){10-12}
           \colhead{ Source }&
           \colhead{ R.A. } &
           \colhead{ Dec. } &
           \colhead{ $R_{\rm int}$ } &
           \colhead{ $F_{\rm int}$ } &
           \colhead{ $F_{\rm int-bg}$ } &
           \colhead{ $R_{\rm int}$ } &
           \colhead{ $F_{\rm int}$ } &
           \colhead{ $F_{\rm int-bg}$ } &
           \colhead{ $R_{\rm int}$ } &
           \colhead{ $F_{\rm int}$ } &
           \colhead{ $F_{\rm int-bg}$ }\\
	   	   \colhead{  } &
           \colhead{ (J2000) }&
           \colhead{ (J2000) }&
	   \colhead{ ($\arcsec$) } &
	   \colhead{ (Jy) } &
	   \colhead{ (Jy) } &
	   \colhead{ ($\arcsec$) } &
	   \colhead{ (Jy) } &
	   \colhead{ (Jy) } &
	   \colhead{ ($\arcsec$) } &
	   \colhead{ (Jy) } &
	   \colhead{ (Jy) } &
}
\startdata
\textbf{Sgr\,D}      &               &               &      &            &      &            &      &            &     \\
\hline
2	&	17 48 33.92	&	-28 02 26.7	&	15	&$>$16.8$\dagger$    & $>$3.93$\dagger$	&\nodata  &\nodata   &\nodata   &	15	& $>$45.3$\dagger$   & $>$28.4$\dagger$		\\
3	&	17 48 41.57	&	-28 08 39.1	&	54	& 587	    & 385		        &\nodata  &\nodata   &\nodata   &	54	& 1571	&	1460				\\
A	&	17 48 35.31	&	-28 00 30.7	&	19	& $<$6.5$\ddagger$	&	\nodata		    &\nodata  &\nodata   &\nodata   &	19	& 39.3	&	23.9			\\
B	&	17 48 41.51	&	-28 02 31.3	&	9	& $<$0.39$\ddagger$	&	\nodata		    &\nodata  &\nodata   &\nodata   &	9	& 6.47	&	4.78				\\
C	&	17 48 43.25	&	-28 01 46.8	&	12	& 15.3	    &	7.99	        &\nodata  &\nodata   &\nodata   &	12	& 74.1	&	72.5				\\
D	&	17 48 48.55	&	-28 01 11.6	&	5	& 2.58	&2.56	        &\nodata  &\nodata   &\nodata   &	8	& 29.2	&	24.6			\\
\hline
\textbf{W42}      &               &               &      &            &      &            &      &            &       \\
\hline
G25.3824	&	18 38 15.38	&	-06 47 52.3	&\nodata	&\nodata   &\nodata   &   8	&862.7	&	797.5	&	9	&1494	&	1326.6	 \\
W42-MME		&	18 38 14.53	&	-06 48 02.3	&\nodata	&\nodata   &\nodata   &   6	&305.7	&	220.5	&	7	&593.5	&	470.5	 \\
1			&	18 38 15.36	&	-06 47 40.8	&\nodata	&\nodata   &\nodata   &   4	&62.7	&	14.7	&	4	&$<$85.3$\ddagger$	&	\nodata
\enddata
\tablecomments{R.A. and Dec. are for the center of apertures used, not the source peaks. $F_{\rm int}$ indicates total flux inside the aperture. $F_{\rm int-bg}$ is for background subtracted flux.}
\tablenotetext{\dagger}{Sgr\,D source 2 is partially off-field in the SOFIA data. The $F_{\rm int}$ and $F_{\rm int-bg}$ values reported are thus lower limits.}
\tablenotetext{\ddagger}{Upper limits values given due to non-detection or for sources that are not sufficiently resolved from background emission.}
\end{deluxetable*}

\begin{deluxetable*}{rcccccccccccc}
\centering
\tabletypesize{\scriptsize}
\tablecolumns{13}
\tablewidth{0pt}
\tablecaption{Spitzer-IRAC Observational Parameters for Sources in Sgr\,D and W42}\label{tb:obspar2}
\tablehead{\colhead{  }&
           \multicolumn{3}{c}{${\rm 3.6\mu{m}}$}&
           \multicolumn{3}{c}{${\rm 4.5\mu{m}}$}&
           \multicolumn{3}{c}{${\rm 5.8\mu{m}}$}&
           \multicolumn{3}{c}{${\rm 8.0\mu{m}}$}\\
           \cmidrule(lr){2-4} \cmidrule(lr){5-7} \cmidrule(lr){8-10} \cmidrule(lr){11-13}
           \colhead{ Source }&
           \colhead{ $R_{\rm int}$ } &
           \colhead{ $F_{\rm int}$ } &
           \colhead{ $F_{\rm int-bg}$ } &
           \colhead{ $R_{\rm int}$ } &
           \colhead{ $F_{\rm int}$ } &
           \colhead{ $F_{\rm int-bg}$ } &
           \colhead{ $R_{\rm int}$ } &
           \colhead{ $F_{\rm int}$ } &
           \colhead{ $F_{\rm int-bg}$ } &
           \colhead{ $R_{\rm int}$ } &
           \colhead{ $F_{\rm int}$ } &
           \colhead{ $F_{\rm int-bg}$ }\\
	   \colhead{  } &
	   \colhead{ ($\arcsec$) } &
	   \colhead{ (Jy) } &
	   \colhead{ (Jy) } &
	   \colhead{ ($\arcsec$) } &
	   \colhead{ (Jy) } &
	   \colhead{ (Jy) } &
	   \colhead{ ($\arcsec$) } &
	   \colhead{ (Jy) } &
	   \colhead{ (Jy) } &
	   \colhead{ ($\arcsec$) } &
	   \colhead{ (Jy) } &
	   \colhead{ (Jy) }
}
\startdata
\textbf{Sgr\,D}      &               &               &      &            &      &            &      &            &      &           &        &      \\
\hline
2	&	18	&0.380	&	0.212	&	18	&0.407	&	0.252	    &	21	&3.51	&	2.00	    &	21	&8.20	&	4.89	\\
3	&	53	&4.22	&	2.16	&	53	&   4.54&	2.67	    &	53	&28.9    &	15.1	    &	53	&78.2	&	46.7	\\
A	&	21	&0.596	&	0.276	&	21	&0.522	&	0.238	    &	24	&4.60	&	1.63	    &  	24	&11.2	&	5.48	\\
B	&	9	&$<$0.051$\dagger$	&\nodata	&	9	&$<$0.060$\dagger$	&\nodata	   &	12	&0.747	&	0.293	    &	12	&1.64	&	0.660	\\
C	&	18	&0.529	&	0.227	&	18	&0.565	&	0.301	    &	21	&3.25	&	1.51	    &	21	&8.18	&	4.30	\\
D	&	6	&0.095	&	0.083	&	6	&0.331	&	0.307	    &	6	&0.820	&	0.658	    &	6	&1.23	&	0.861	\\
\hline
\textbf{W42}      &               &               &      &            &      &            &      &            &      &           &        &      \\
\hline
G25.3824&	5	&1.06	&	0.738	&	5	&1.27	&	0.922	&	7	&8.81	&	5.40	&	\nodata	& sat.	&	sat.	\\
W42-MME	&	3	&0.23	&	0.140	&	4	&0.819	&	0.654	&	4	&3.67	&	2.46	&	\nodata	& sat.	&	sat.	\\
1	    &	5	&0.22	&	0.054	&	4	&0.183	&	0.043	&	5	&1.55	&	0.364	&	\nodata	& sat.	&	sat.
\enddata
\tablecomments{\footnotesize $F_{\rm int}$ indicates total flux inside the aperture. $F_{\rm int-bg}$ is for background subtracted flux. `sat.' means the source is saturated at that wavelength, and thus no accurate flux can be measured.}
\tablenotetext{\dagger}{The $F_{\rm int}$ value is used as the upper limit since the source is difficult to distinguish above the background.}
\end{deluxetable*}

\begin{deluxetable*}{rrrrrrr}
\centering
\tabletypesize{\scriptsize}
\tablecolumns{7}
\tablewidth{0pt}
\tablecaption{Herschel-PACS Observational Parameters for Sources in Sgr\,D and W42}\label{tb:obspar3}
\tablehead{\colhead{  }&
           \multicolumn{3}{c}{${\rm 70\mu{m}}$}&
           \multicolumn{3}{c}{${\rm 160\mu{m}}$}\\
           \cmidrule(lr){2-4} \cmidrule(lr){5-7} 
           \colhead{ Source }&
           \colhead{ $R_{\rm int}$ } &
           \colhead{ $F_{\rm int}$ } &
           \colhead{ $F_{\rm int-bg}$ } &
           \colhead{ $R_{\rm int}$ } &
           \colhead{ $F_{\rm int}$ } &
           \colhead{ $F_{\rm int-bg}$ }\\
	   	   \colhead{  } &
	   \colhead{ ($\arcsec$) } &
	   \colhead{ (Jy) } &
	   \colhead{ (Jy) } &
	   \colhead{ ($\arcsec$) } &
	   \colhead{ (Jy) } &
	   \colhead{ (Jy) } 
}
\startdata
\textbf{Sgr\,D}     &           &   &           &           &   &      \\
\hline
2	                &	26		&382	&	241 &	26		&844	&  157	\\
3	                &	48		&3160	&	2330	&	64		&6230	&  2695	\\
A	                &	38		&685	&	244	&	38		&$<$1648$\dagger$	&	\nodata	\\
B               	&	16		&97.2	&	22.7	&	16		&$<$265$\dagger$	    &	\nodata	\\
C	                &	19	    &$<$612	&	UR		&	19	&$<$733	        &	UR		\\
D	                &	16		&163	    &	96.2	&	16		&791	&	97.2 \\
\hline
\textbf{W42}       &            &   &           &           &   &    \\
\hline
G25.3824	       &	13	    &	$<$2207$\dagger$ &\nodata	&	13	&	$<$840$\dagger$ &\nodata	\\
W42-MME		       &	10	    &	$<$1040$\dagger$ &\nodata	&	10	&	$<$330$\dagger$ &\nodata	\\
1		     	   &	10	    &	$<$637$\dagger$  &\nodata	&	10	&	$<$318$\dagger$ &\nodata
\enddata
\tablecomments{$F_{\rm int}$ indicates total flux inside the aperture. $F_{\rm int-bg}$ is for background subtracted flux. `UR' means the source is not sufficiently resolved from the much brighter Source 3, and thus flux values can only be considered upper limits.}
\tablenotetext{\dagger}{The $F_{\rm int}$ value is used as the upper limit since the source is difficult to distinguish above the background.}
\end{deluxetable*}

\begin{figure*}[tb!]
\epsscale{1.00}
\plotone{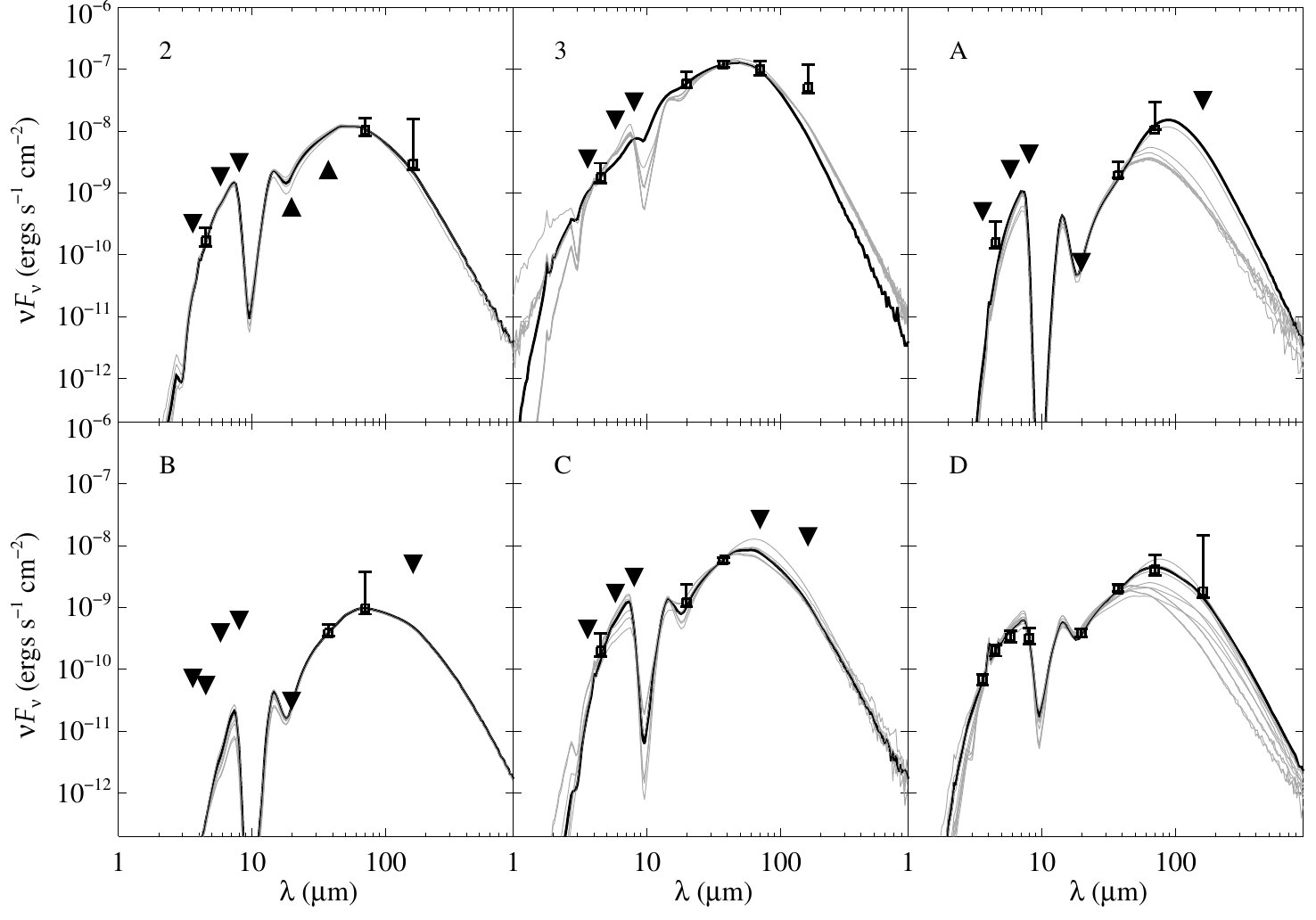}
\caption{SED fitting with ZT model for compact sources in Sgr\,D. Black lines are the best fit model to the SEDs, and the system of gray lines are the remaining fits in the group of best fits (from Table~\ref{tb:sed}). Upside-down triangles are data that are used as upper limits in the SED fits. \label{fig:SgrD_sed}}
\end{figure*}

\begin{figure*}[tb!]
\epsscale{1.00}
\plotone{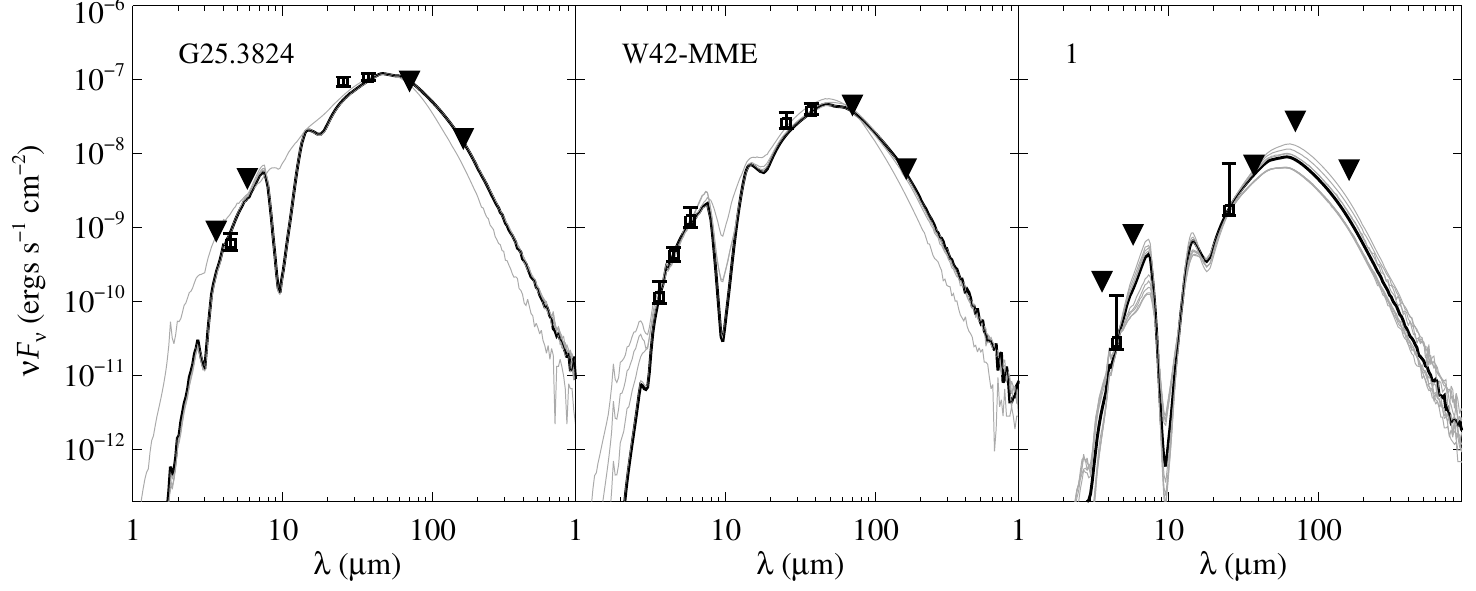}
\caption{SED fitting with ZT model for compact sources in W42. See Figure~\ref{fig:SgrD_sed} for details. \label{fig:W42_sed}}
\end{figure*}

\begin{deluxetable*}{rccccrclrclcl}
\tabletypesize{\small}
\tablecolumns{12}
\tablewidth{0pt}
\tablecaption{SED Fitting Parameters for Sources in Sgr\,D and W42}\label{tb:sed}
\tablehead{\colhead{   Source   }                                              &
           \colhead{  $L_{\rm obs}$   } &
           \colhead{  $L_{\rm tot}$   } &
           \colhead{ $A_v$ } &
           \colhead{  $M_{\rm star}$  } &
           \multicolumn{3}{c}{$A_v$ Range}&
           \multicolumn{3}{c}{$M_{\rm star}$ Range}&
           \colhead{ Best }&
           \colhead{Notes}\\
	   \colhead{        } &
	   \colhead{ ($\times 10^3 L_{\sun}$) } &
	   \colhead{ ($\times 10^3 L_{\sun}$) } &
	   \colhead{ (mag.) } &
	   \colhead{ ($M_{\sun}$) } &
       \multicolumn{3}{c}{(mag.)}&
       \multicolumn{3}{c}{($M_{\sun}$)}&
       \colhead{  Models   } &
       \colhead{   }
}
\startdata
\textbf{Sgr\,D}  &       &        &      &     &     &          &     &  &      &   &     \\
\hline
2 &      3.01 &     10.18 &     72.1 &      8.0 &  55.3 & - & 106.0 &   8.0 & - &  16.0 &  5  & MYSO\\
3 &     33.91 &    111.64 &     26.5 &     16.0 &   1.7 & - &  32.7 &  16.0 & - &  32.0 & 11  & MYSO$\dagger$ \\
A &      2.74 &      4.95 &    264.9 &      2.0 & 257.0 & - & 293.0 &   2.0 & - &  32.0 &  8 & \\
B &      0.24 &      0.35 &     76.3 &      2.0 &  60.4 & - &  80.5 &   2.0 & - &   2.0 &  7 & \\
C &      1.98 &     11.66 &     92.7 &      8.0 &  47.8 & - & 162.0 &   8.0 & - &  24.0 &  7 & MYSO\\
D &      1.10 &      1.45 &     74.2 &      4.0 &  58.7 & - & 109.0 &   2.0 & - &   8.0 & 11 & \\
\hline
\textbf{W42}  &       &        &      &     &     &          &     &  &      &   &     \\
\hline
G25.3824 &     24.54 &     81.95 &     40.2 &     24.0 &  26.5 & - &  44.4 &  24.0 & - &  24.0 &  7 & MYSO \\
W42-MME &      9.14 &    146.68 &     82.2 &     32.0 &  23.8 & - &  82.2 &  16.0 & - &  32.0 &  5 & MYSO  \\
1   &      1.63 &     11.66 &     98.0 &      8.0 &  58.3 & - & 159.0 &   8.0 & - &  24.0 & 10 & MYSO 
\enddata
\tablecomments{\footnotesize A ``MYSO'' in the right column denotes a MYSO candidate. A MYSO candidate has values for both $M_{\rm star}$ and its whole $M_{\rm star}$ range greater than 8\,$M_{\sun}$.}
\tablenotetext{\dagger}{\footnotesize No SED fits can be found for source Sgr\,D source 3 that go through the 160\,$\mu$m data point, indicating a possible excess of a colder environmental dust present.}
\end{deluxetable*}

In order to perform photometry on mid-infrared point sources, we employed the aperture photometry program \textit{aper.pro}, which is part of the IDL DAOPHOT package available in The IDL Astronomy User's Library (http://idlastro.gsfc.nasa.gov). As was done in \citetalias{2019ApJ...873...51L}, we measured flux densities for all compact sources and sub-regions that could be identified in the SOFIA 20\,$\mu$m (or 25\,$\mu$m) and 37 \,$\mu$m data for Sgr\,D and W42. We additionally downloaded Spitzer-IRAC (i.e., 3.6, 4.5, 5.8, 8.0\,$\mu$m) imaging data and Herschel-PACS (i.e., 70 and 160\,$\mu$m) imaging data from their respective online archives and measured fluxes for these same sources at all wavelengths.  Tables \ref{tb:obspar1}-\ref{tb:obspar3} contains the information regarding the position, radius employed for aperture photometry, and background subtracted flux densities measured at all wavelengths for all of these sources. We employed the same optimal extraction technique as in \citetalias{2019ApJ...873...51L} to find the optimal aperture to use for photometry. Background subtraction was also performed in the same way as \citetalias{2019ApJ...873...51L} (i.e. using background statistics from an annulus outside the optimal extraction radius which had the least environmental contamination). For sources in W42, we could not determine source fluxes at 8\,$\mu$m due to the Spitzer image being saturated in these areas. For source D in Sgr\,D, our on-source field of view did not cover this source,  however we were chopping east-west which picked up source D in the eastern chop reference beam, and therefore it shows up as a negative source in our data. Though it is displaced to the west by the distance of the chop throw (180$\arcsec$) and negative, accurate photometry could still be performed on the source.

We also used the same methodology set out in \citetalias{2019ApJ...873...51L} to determine which Spitzer-IRAC data points could be considered nominal data points and which should be considered only upper limits based upon potential contamination from PAH and/or shock-excited H$_2$ emission. As seen in Figure~\ref{fig:cc}, all but two sources are considered ``PAH emission dominant'' meaning that for those sources we consider their 3.6, 5.8, and 8.0\,$\mu$m Spitzer data to be upper limits only. As we did in our previous papers, the Herschel 70 and 160\,$\mu$m data are set to be upper limits for about half of the observations due to the coarser spatial resolution ($\sim$10$\arcsec$) of the data and the high likelihood that the photometry is contaminated by emission from adjacent sources or the extended dusty environments in and around Sgr\,D and W42.  

We set the upper error bar on our photometry as the subtracted background flux value (since background subtraction can be highly variable but never larger than the amount being subtracted), and the lower error bar values for all sources come from the average total photometric error at each wavelength (as discussed in Section~2 and \citetalias{2019ApJ...873...51L}) which are set to be the estimated photometric errors of 20\%, 15\%, and 10\% for 4.5, 20/25, and 37\,$\mu$m bands, respectively. We assume that the photometric errors of the Spitzer-IRAC 3.6, 5.8, and 8.0\,$\micron$ fluxes are 20\% for the sources that are not contaminated by PAH features. The lower error bars of the 70 and 160\,$\micron$ data points are assumed to be 40$\%$ and 30$\%$, respectively, adopting the most conservative (largest) uncertainties of the Herschel compact source catalog \citep{2016AA...591A.149M,2017MNRAS.471..100E}.

We used these photometry data to create the near-infrared to far-infrared SEDs of the identified sources and fit them with SED models of MYSOs \citep[a.k.a. ``ZT Models'';][]{2011ApJ...733...55Z} following the same procedures as described in \citetalias{2019ApJ...873...51L}. As we did in our previous papers, we used the non-background subtracted fluxes for the 3.6, 5.8, and 8.0\,$\mu$m Spitzer-IRAC data in the PAH contaminated sources when fitting the models.  We show the ZT MYSO SED model fits as the solid lines (black for the best model fit, and gray for the rest in the group of best fit models) on top of the derived photometry points for each individual source in Sgr\,D (Figure~\ref{fig:SgrD_sed}) and W42 (Figure~\ref{fig:W42_sed}). Table~\ref{tb:sed} lists the physical properties of the MYSO SED model fits for each source. The observed bolometric luminosities, $L_{\rm obs}$, of the best fit models are presented in column~2 and the true total bolometric luminosities, $L_{\rm tot}$ (i.e. corrected for the foreground extinction and outflow viewing angles), in column~3. The extinction and the stellar mass of the best models are listed in column~4 and 5, respectively. The column~6 and 7 present the ranges of the foreground extinction and stellar masses derived from the models in the group of best model fits, and the number of best model fits is given in column~8. Column~9 shows the identification of the individual sources based on the previous studies as well as our criteria of MYSOs and possible MYSOs (``pMYSOs'') set in \citetalias{2019ApJ...873...51L}. To summarize, the conditions for a source to be considered a MYSO is that it must 1) have an SED reasonably fit by the MYSO models, 2) have a M$_{\rm star}\ge8\,$M$_{\sun}$ for the best model fit model, and 3) have M$_{\rm star}\ge8\,$M$_{\sun}$ for the range of $M_{\rm star}$ of the group of best fit models. A pMYSO fulfills only the first two of these criteria.
 
\section{S\lowercase{gr}\,D}\label{sec:SgrD}

\begin{figure}[htb!]
\begin{center}
\includegraphics[width=3.3in]{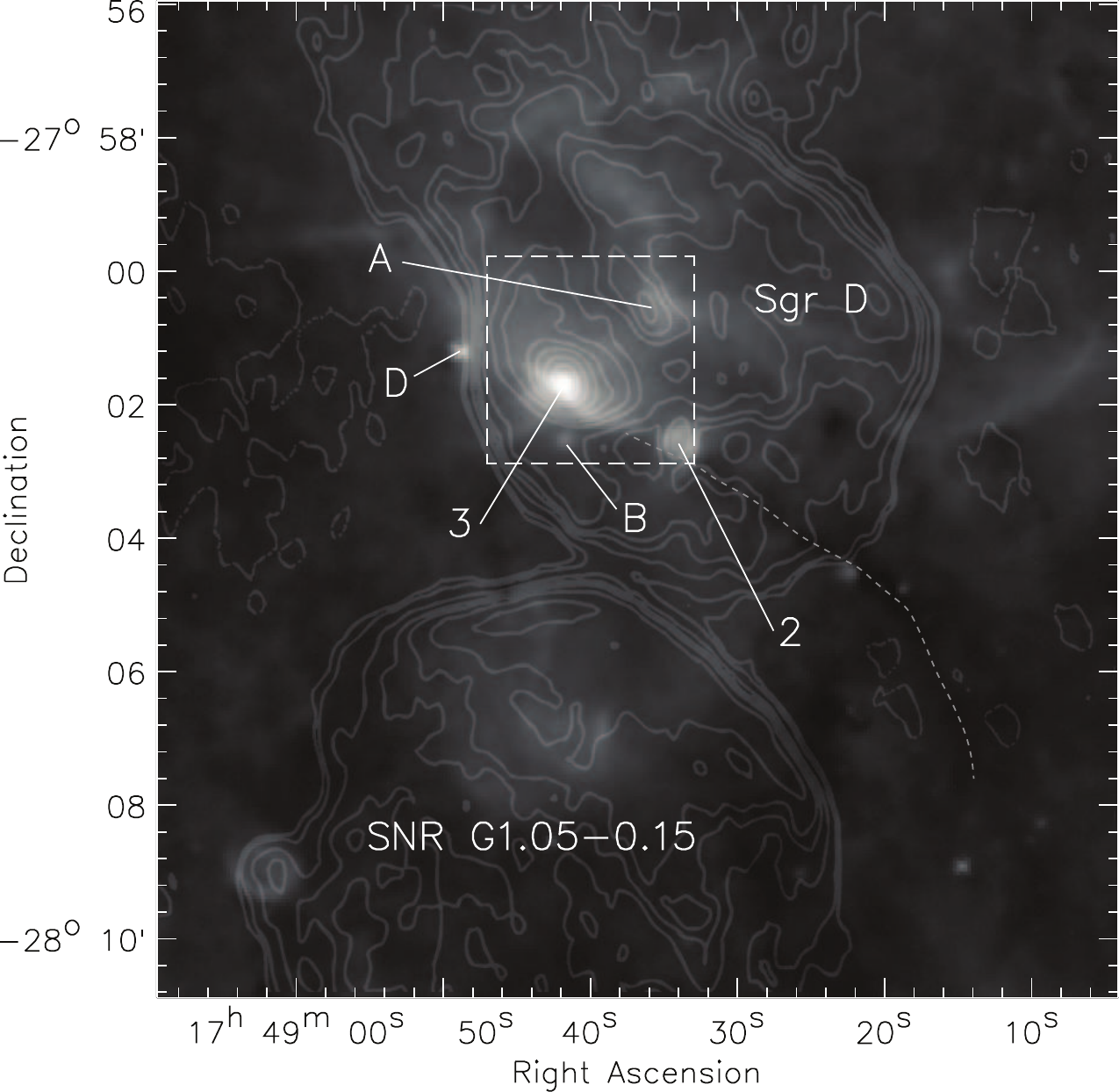}\\
\end{center}
\caption{\footnotesize Sgr D and its neighboring supernova remnant G1.05-0.15 as seen by Herschel-PACS at 70\,$\mu$m overlaid with the 18\,cm radio continuum contours of \citet{1998ApJ...493..274M}. Previously known radio sources 2 and 3 are labeled, as well as infrared sources A, B, and D identified in this work. The white dashed box shows the area covered by the SOFIA observations in Figure\,\ref{fig:SgrD2}. The dotted white line delineates a dark filament.}
\label{fig:SgrD1}
\end{figure}

The 6$\farcm$6-diameter \ion{H}{2} region of Sgr\,D appears fairly circular in cm radio continuum emission \citep[especially at longer radio wavelengths like 18\,cm;][]{2019MNRAS.482.5349R} and lies just north-west of a supernova remnant, G1.05-0.15, which has a similar size (Figure~\ref{fig:SgrD1}). Being a source near the Galactic Center in projection, it was assumed after initial observations \citep[e.g.,][]{1973AA....22..413K} to be at a similar distance from the Sun as Sgr~A*. More recent observations have argued for distances farther than the Galactic Center \citep{1998ApJ...493..274M} as well as closer \citep{1999ApJ...512..237B}. However, using trigonometric parallax measurements of the 22\,GHz water maser emission in Sgr\,D \citep{2017PASJ...69...64S}, the distance has recently been accurately determined, and indeed has been shown to be a lot closer, situated only 2.36\,kpc away. That puts the physical diameter of the \ion{H}{2} region at $\sim$4.5\,pc (rather than $\sim$15\,pc), and using the calculations we will discuss in later in Section~\ref{sec:lyman}, reduces the Lyman continuum photon rate of the entire region from log$N_{LyC} = 50.52$ photon/s to just log$N_{LyC} = 49.37$ photon/s. This is the equivalent to a single O6V star \citep{1973AJ.....78..929P}, and disqualifies Sgr\,D from being a G\ion{H}{2} based upon its Lyman continuum photon rate.

Looking to the Spitzer-IRAC images and the MSX 22\,$\mu$m image shown in \citet{2004MNRAS.355..899C}, most of the 6$\farcm$6-diameter \ion{H}{2} region is not readily apparent in the near and mid-infrared. Even in the Herschel-PACS and SPIRE data, which covers wavelengths with the most infrared emission in the area, the infrared morphology does not resemble the overall radio morphology very well (Figure~\ref{fig:SgrD1}). The fact that most of the radio \ion{H}{2} region is not infrared-bright could indicate that it may have never been in a giant molecular cloud, or it may be old enough that the molecular cloud around could have dissipated. 

\begin{figure*}[tb!]
\epsscale{1.15}
\plotone{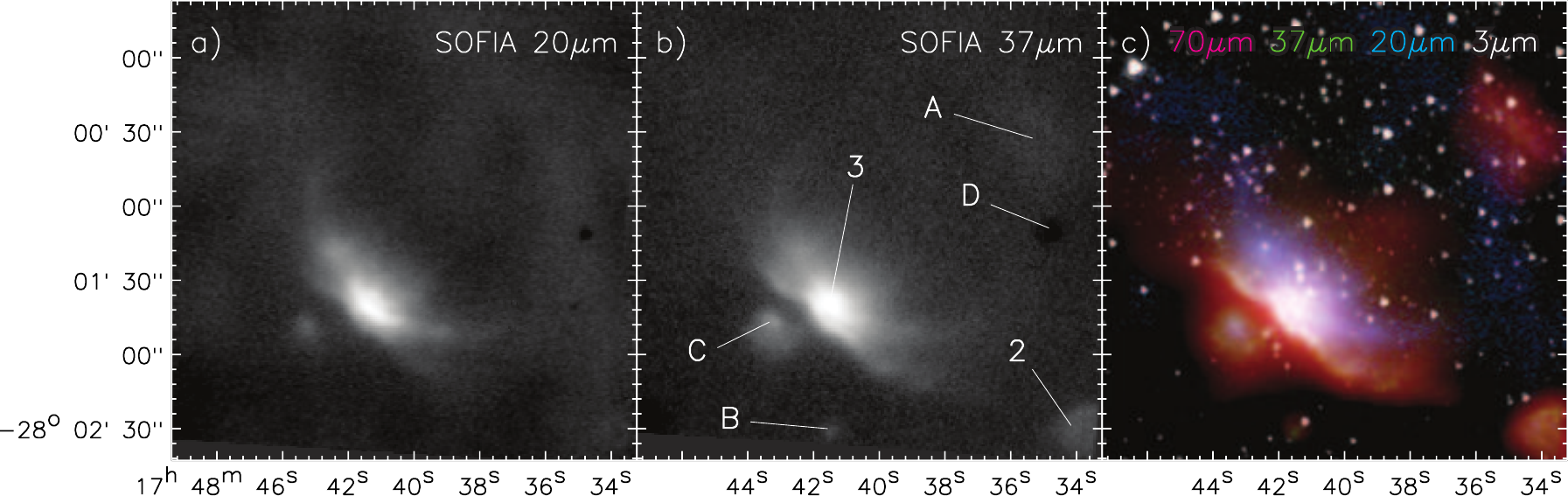}
\caption{Images of the brightest infrared-emitting sources within the Sgr D radio continuum region at a) SOFIA 20\,$\mu$m, and b) SOFIA 37\,$\mu$m. Panel c) is a 4-color image of the same region made with Herschel 70\,$\mu$m (red), SOFIA 37\,$\mu$m (green), SOFIA 20\,$\mu$m (blue), and Spitzer 3.6\,$\mu$m (white; stars) data. Previously known radio sources 2 and 3 are labeled, as well as infrared sources A, B, C, and D identified in this work. Source D is seen as a negative source, but lies just off the SOFIA field to the east (see Figure~\ref{fig:SgrD1}).  \label{fig:SgrD2}}
\end{figure*}

Our SOFIA 20 an 37\,$\mu$m images show the area is dominated in the mid-infrared by a small and bright region (15$\arcsec\times$40$\arcsec$) on the southeastern side of the larger, circular \ion{H}{2} region (Figure~\ref{fig:SgrD2}).  The 6 and 18\,cm radio continuum observations of \citet{1998ApJ...493..274M} show a bright radio peak at this location (named source 3), which they state has the ionizing equivalent of an O5.5 ZAMS star. However, they assume a distance to Sgr\,D of 8.5\,kpc, so when we recalculate this using the method of Section~\ref{sec:lyman} with the new distance of 2.36\,kpc, we derive an ionizing flux equivalent to an O8 ZAMS star \citep[log$N_{LyC}$$ = 48.36$\,photons/s;][]{1973AJ.....78..929P}. \citet{1998ApJ...493..274M} also identify another compact radio source that lies within the diffuse spherical \ion{H}{2} region of Sgr\,D which they label source 2 (see Figure~\ref{fig:SgrD1}), which we calculate has the radio flux equivalent of a B0.5 ZAMS star. We also partially see this source on the edge of the field of our SOFIA data (Figure~\ref{fig:SgrD2}). In addition to these two previously identified radio sources, there are four smaller, infrared sources within the larger radio \ion{H}{2} region of Sgr\,D, which we label A through D in order of right ascension in Figure~\ref{fig:SgrD2}. 

The elongated infrared structure of radio source 3 has a sharp edge to the southeast, indicating a dense structure present, and infrared source C appears as a fainter region of emission $\sim$10$\arcsec$ beyond the sharp edge of source 3 to the southeast. \citet{1989IAUS..136..205O} suggested that source 3 is a blister-type of \ion{H}{2} region seen almost edge-on on the rim of a dark molecular cloud. However, looking to the Herschel 70\,$\mu$m image of the region shows a dark filament running from this source towards the southeast (Figure~\ref{fig:SgrD1}), and this filament can be seen in emission in the Herschel 250, 350, and 500\,$\mu$m images. Therefore source 3 likely formed on the edge of this narrow mid-infrared-dark filament instead of a cloud. In fact, the dark filament runs parallel to the outer contours of the supernova remnant G1.05-0.15 (Figure~\ref{fig:SgrD1}), indicating that either the filament is influencing the shape of the remnant or that the filament is caused by the sweeping up of material from the expanding supernova remnant. Interestingly, sources 2, 3, and D all lie along a line which meets up with the dark filament to the southwest. This indicates that these sources were likely formed out of that filament, which was perhaps induced by the collision of the Sgr\,D \ion{H}{2} region with the SNR. 

In both the SOFIA 20 and 37\,$\mu$m images the elongated emission of source 3 branches into an X-shape, and curls away from what appears to be a dark lane which is likely the continuation of the filament. These SOFIA images are reminiscent of an almost edge-on flared disk, where the dark lane separating the brighter and fainter regions of emission is the optically thick torus/disk mid-plane, and the brighter northwest infrared emission of source 3 curls away from the dark mid-plane because it is coming from the flared disk surface, and the fainter southeastern emission from source C would be from the other, more obscured side of the disk. Therefore the infrared emission from source 3 and C could be coming from the same ionizing object that resides in a disk within the filament, or the filament is geometrically thin enough that the morphology seen in the infrared mimics that of a disk. 
 
Based upon the ZT Model fits to the source SEDs,  we find source 3 is likely to be a MYSO with a mass of 16-32\,$M_{\sun}$ (Table~\ref{tb:sed}). However, it is the only source that is not well-fit by the SED models, since no fit could be found that goes close to or through the Herschel 160\,$\mu$m data point. As this source is embedded within a larger cold filament, there may be an excess of cold dust in its immediate environment (or seen in projection at our viewing angle) that is not well fit by the core-accretion SED models. That being said, even with SED models that underfit the 160\,$\mu$m emission, this source appears to be the most massive YSO of all the Sgr\,D sources identified in the SOFIA data. The range of infrared-derived mass would be the equivalent of a B0.5-O7 ZAMS star \citep{2000AJ....119.1860B}, which at the maximum extent of this range is consistent with the previously mentioned spectral type of O8 ZAMS star derived from the radio data.  

Source A is near the center of the 6$\farcm$6-diameter \ion{H}{2} region and it associated with a weak cm radio peak (Figure~\ref{fig:SgrD1}). This source is a very weak and diffuse patch of emission in the mid-infrared with no real discernible peak, and not extremely prominent even in the Herschel 70\,$\mu$m data. Our best fit SED to the infrared data yield a estimated stellar mass of only 2.0\,$M_{\sun}$. There is a considerable drop in goodness of fit (as seen in Figure~\ref{fig:SgrD_sed} with several models underfitting the 70\,$\mu$m data point), and those models have masses as high as 32\,$M_{\sun}$. Given the best fit model is of a low-mass star, we do not consider this source a MYSO. 
 
Source B is not seen in the SOFIA 20\,$\mu$m images and is faint in the Spitzer-IRAC images (Figure~\ref{fig:SgrD2}), but can be seen clearly in the Herschel 70\,$\mu$m image (Figure~\ref{fig:SgrD1}). Our SED modeling of this source show it to be a low mass YSO of 2\,$M_{\sun}$, however 6 of the 9 data points being fit are upper limits, and thus the model fitting is not well-constrained.

As stated earlier, although source C might be emission associated with source 3, we modeled it as an independent source in the event that it indeed is. It is faint in the SOFIA 20\,$\mu$m image and is brighter and more extended in the SOFIA 37\,$\mu$m image (Figure~\ref{fig:SgrD2}). It is seen as a peakless but relatively bright emission region extended off of source 3 in the Herschel 70\,$\mu$m image. Our SED modeling shows that, if it indeed is an independent source, it would be a MYSO candidate with a derived mass of 8\,$M_{\sun}$.

The location of source D is best seen in the Herschel 70\,$\mu$m data in Figure~\ref{fig:SgrD1}, where it appears as a bright point source. As stated in Section~\ref{sec:data}, Source D was not on the field of our SOFIA data, however we were chopping east-west and source D was in our eastern chop reference beam, and therefore shows up as a negative source in our data in Figure~\ref{fig:SgrD2}. Our SED modeling show it to be an intermediate mass YSO with a best fit mass of 4\,$M_{\sun}$, though model fits do go as high as $8\,M_{\sun}$. 

Though source 2 lies partially off the SOFIA field at 20 and 37\,$\mu$m, we derived SED fits for the \text{Spitzer} and Herschel data for the source and used the SOFIA photometry of the partially seen source as lower limits. This yielded mass estimates of 16-32\,$M_{\sun}$ for the source. This range is consistent with the radio-derived mass estimate of $\sim$16\,$M_{\sun}$ (i.e., a B0.5 ZAMS star).

In summary, the majority of the 6$\farcm$6-diameter radio-emitting region of Sgr\,D does not have any significant infrared emission and thus appears to be the location of only a low level of newly forming stars. Our SED modeling of the infrared sources in Sgr\,D show that the star formation in the region is dominated by one source, the 16\,$M_{\sun}$  source 3, and that there are likely two (and no more than four) other MYSOs present. Based upon these results, it seems that this region is predominantly powered by a single source (source 3) but, as we will discuss in more detail in Section~\ref{sec:reject}, this source is of relatively modest size compared to the sources powering the G\ion{H}{2} regions previously studied in our survey. 

\section{W42}\label{sec:W42}

\begin{figure*}[tb!]
\epsscale{1.15}
\plotone{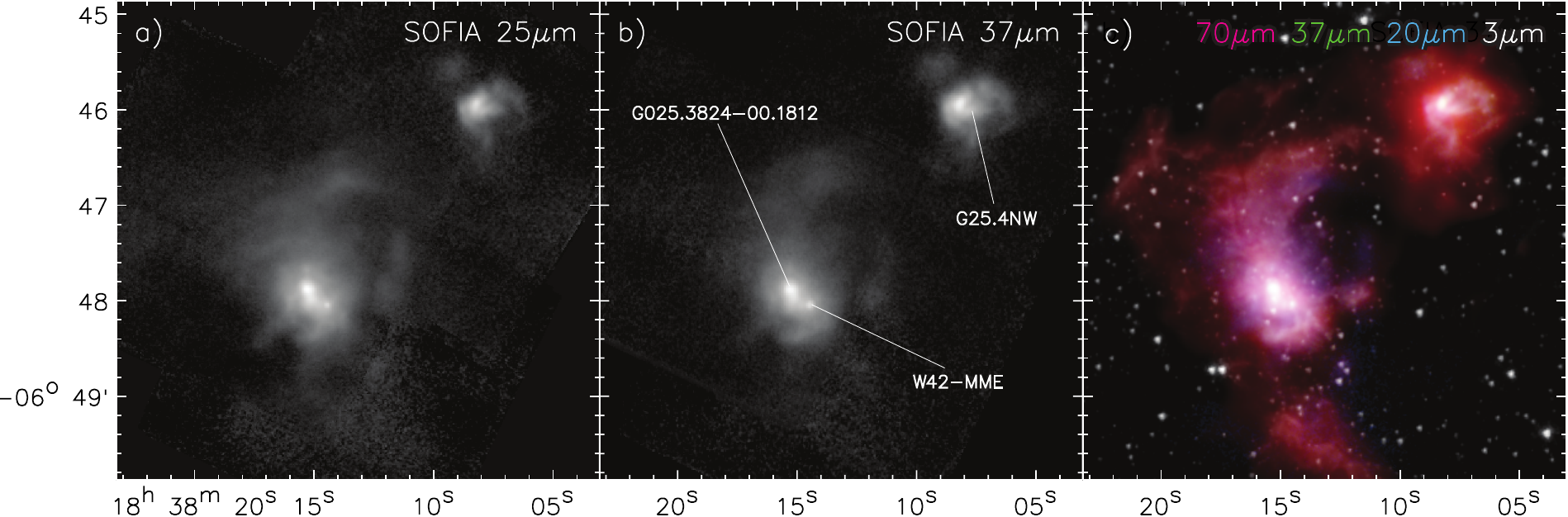}
\caption{SOFIA images of W42 and G25.4NW at a) SOFIA 25\,$\mu$m, and b) SOFIA 37\,$\mu$m. Panel c) is a 4-color image of the same region made with Herschel 70\,$\mu$m (red), SOFIA 37\,$\mu$m (green), SOFIA 25\,$\mu$m (blue), and Spitzer 3.6\,$\mu$m (white; stars) data. G25.4NW lies at a different distance than the rest of W42 and is thus not related. The two brightest peaks seen in the mid-infrared are labeled and coincident with the compact radio source G025.3824-00.1812 and the methanol maser source W42-MME. \label{fig:W42_1}}
\end{figure*} 

W42, also known as G25.38–0.18, has cm radio continuum emission that is described as having a core-halo structure with a diameter of $\sim$3$\arcmin$ \citep{1993ApJ...418..368G}. Observations in $^{13}$CO towards W42 show a line velocity of  58-69\,km\,s$^{-1}$ \citep[e.g.,][]{2009ApJ...690..706A,2013RAA....13..935A}, leading to a near kinematic distance of $\sim$3.7\,kpc and a far distance of $\sim$11.2\,kpc. The far kinematic distance was assumed by \citet{2004MNRAS.355..899C}. However, \citet{2000AJ....119.1860B} were able to fit the near-infrared spectra of the brightest star near the center of W42 (labeled as W42\,\#1) with a zero-age main sequence O5–O6.5 spectral type, which led them to the derivation of a spectrophotometric distance of $\sim$2.2\,kpc. Further evidence that this closer value is more accurate comes from \citet{2018ApJ...864..136B}, who adopt this 2.2\,kpc value because even the near kinematic value of $\sim$3.7\,kpc would make their measured luminosity for the region inconsistent with the cataloged massive stellar population. \citet{2000AJ....119.1860B} give no error on their 2.2\,kpc value but state that the uncertainty in the distance estimate is dominated by the luminosity class assumed and the scatter in the intrinsic brightness of the O stars. They do, however, quote a distance of $2.6^{+1.0}_{-0.7}$\,kpc using such errors under the assumption that the stars are main sequence dwarf stars, and $3.4^{+1.2}_{-0.9}$\,kpc for main sequence giant stars. The assumption of a ZAMS type is more reasonable, and assuming a comparable level of errors for ZAMS stars (i.e. $+37$\% and $-27$\%), we will assume a distance and errors of $2.2^{+0.8}_{-0.6}$\,kpc in this work. \citet{2011MNRAS.411..705M} also derived distances to W42 spectrophotometrically and got a value of 2.67$\pm$1.40\,kpc, which is consistent with the \citet{2000AJ....119.1860B} value and our assumed errors.

At this new distance, the derived log$N_{LyC}$ value for the entire W42 \ion{H}{2} region is 49.44\,photons/s (see Section~\ref{sec:lyman}), which for comparison is 79\% the value for Orion. This value is consistent with the Lyman continuum photon rate of a single $\sim44\,M_{\sun}$ O5.5 ZAMS star \citep{1973AJ.....78..929P}, and thus the centrally-located star W42\,\#1 found by \citet{2000AJ....119.1860B} is thought to be almost fully responsible for ionizing the entire W42 \ion{H}{2} region. 

The MSX 22\,$\mu$m image shown in \citet{2004MNRAS.355..899C} displays two sources, with the brighter, more extended source to the southeast being W42, and the second to the northwest being a source called G25.4NW. While G25.4NW also displays cm continuum radio emission and is only 2.6$\arcmin$ from the peak of W42, $^{13}$CO line profiles show that it is at a very different velocity from W42 \citep[e.g.,][]{2013RAA....13..935A}, and thus the two sources lie at different distances and are not physically associated. In our SOFIA data we do see both the W42 \ion{H}{2} region and G25.4NW at both 25 and 37\,$\mu$m (Figure~\ref{fig:W42_1}), however we will not discuss G25.4NW any further here.

Looking at the large-scale structure of the region, the MSX 22\,$\mu$m image of W42 \citep{2004MNRAS.355..899C} reveals a source with a bright central region with extended emission elongated ($\sim$3$\arcmin$) north-south. Spitzer-IRAC images of W42 revealed fainter dust emission extending perpendicular to, and extending much farther ($r\sim5\arcmin$) than, what was seen by MSX. \citet{2015ApJ...811...79D} believe that this fainter emission shows a bipolar structure, roughly east-west, with the brighter, centrally-located, north-south elongated infrared emission being central the waist of the overall bipolar structure. Our SOFIA observations at 25 and 37\,$\mu$m only detect the extended emission of W42 out to about the same extent as that seen in the MSX 22\,$\mu$m image (Figure~\ref{fig:W42_1}), however with much better resolution ($\sim$3$\arcsec$ versus $\sim$18$\arcsec$). 

\begin{figure}[tb!]
\begin{center}
\includegraphics[width=3.3in]{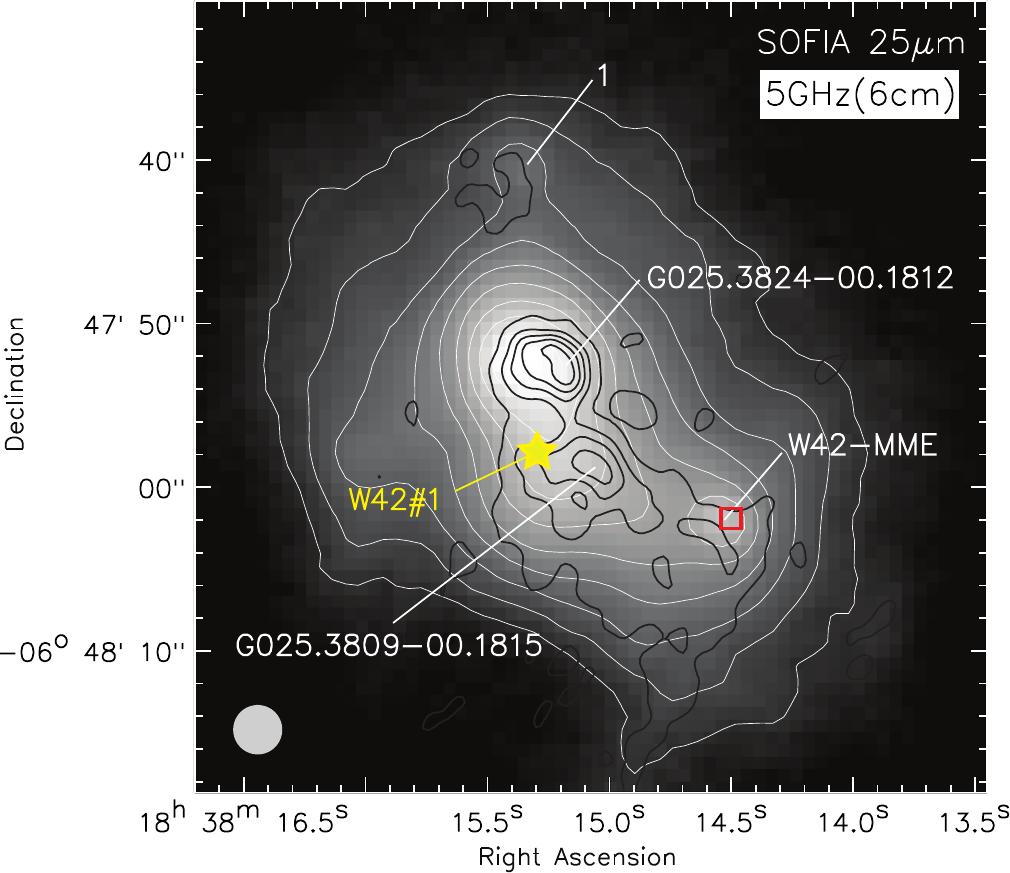}\\
\end{center}
\caption{\footnotesize The inner area of W42. The grayscale image and white contours are the SOFIA 25\,$\mu$m data overlaid with the 5\,GHz (6\,cm) radio continuum contours (black) from the CORNISH survey \citep{2013ApJS..205....1P}. Radio sources G025.3824-00.1812 and G025.3809-00.1815 are identified, as well as the infrared peak of W42-MME. Newly identified source 1 is also labeled. The yellow star marks the location of the revealed O5-O6.5V star W42\,\#1. The red square shows the location of the methanol maser emission detected by \citet{2012AN....333..634S} that is coincident with W42-MME. The gray dot in the lower left corner gives the angular resolution of the 25\,$\mu$m image. }
\label{fig:W42_2}
\end{figure}

On smaller scales, looking to the SOFIA images, we find the central region of W42 contains multiple peaks (Figure~\ref{fig:W42_2}). The brightest peak in the 25 and 37\,$\mu$m data corresponds to a UC\ion{H}{2} region named G025.3824-00.1812 (labeled `G25.3824', for short, in Table~\ref{tb:sed} and Figures~\ref{fig:cc} and \ref{fig:W42_sed}) resolved by the 5\,GHz (6\,cm) CORNISH Survey \citep{2013ApJS..205....1P}. This also corresponds to the only peak in the extended emission of W42 seen in the Herschel 70\,$\mu$m image, as well as the brightest peak seen in the Spitzer-IRAC images. \citet{2015ApJ...811...79D} obtained high spatial resolution near infrared images of this source (0.027$\arcsec$) that resolve the source into five point sources, but they were not able to characterize the sources individually. Based on the radio flux of G025.3824-00.1812 \citep[$S_{5\,\mathrm{GHz}}$$=200.13$\,mJy;][]{2013ApJS..205....1P}, we calculate an $N_{LyC}$ for this region alone and find that it is ionized by the equivalent of a single B0.5V star, which would have a mass of $M\sim16M_{\sun}$. Our infrared-derived mass from the best fits of the SED modeling yields a moderately larger stellar mass of $24\,M_{\sun}$ (Table~\ref{tb:sed}). 

The next brightest peak in the SOFIA data is the southwestern peak that is associated with the location of a 6.7 GHz methanol maser source \citep{2012AN....333..634S}. \citet{2015ApJ...803..100D} detect an infrared source here at wavelengths longer than 2.2\,$\mu$m (it is also prominent in the Spitzer-IRAC images) which they call W42-MME. They classify this source as a deeply embedded massive YSO ($M\sim14\,M_{\sun}$), which is driving a parsec-scale H$_2$ outflow.  Based upon our SOFIA data and SED models we confirm the nature of this source as being a MYSO, with a best fit mass of $32\,M_{\sun}$ and with the lower limit of the mass range of the best fit models ($16\,M_{\sun}$; Table~\ref{tb:sed}) being consistent with the previously derived mass estimate.

There is no peak in the dust emission in the SOFIA data at the location of the central O5-O6.5 star, W42\,\#1, identified by \citet{2000AJ....119.1860B} though the star itself is easily seen in the Spitzer-IRAC images. This, along with the fact that the source was not so deeply embedded as to allow its spectroscopic identification, may indicate that W42\,\#1 has evolved far enough to have expelled its natal envelope. 

There is a region of emission seen by SOFIA extending southwest from G025.3824-00.1812 $\sim$5$\arcsec$ west of W42\,\#1, that corresponds to the radio region G025.3809-00.1815 \citep[CORNISH Survey;][]{2013ApJS..205....1P}.  Based on the radio flux ($S_{5\,
\mathrm{GHz}}=460.83$\,mJy), we calculate an $N_{LyC}$ for this region alone and find that it is ionized by the equivalent of a single B0V star. It is unclear how much of the radio flux of G025.3824-00.1812 is due to self-luminance or ionization by W42\,\#1 nearby. This goes for 025.3809-00.1815 as well, given that it too is only $\sim$5$\arcsec$ away from W42\,\#1. We cannot isolate the emission from 025.3809-00.1815 well enough to get accurate fluxes for our SED modeling. As for the near-infrared emission associated with W42\,\#1, \citet{2015ApJ...811...79D} resolve it into three point sources, but do not have sufficient data to classify them individually. One would assume that the brightest of the three objects is likely to be the $\sim44\,M_{\sun}$ O5-O6.5 star identified spectroscopically by \citet{2000AJ....119.1860B}.

There is one final peak seen in the SOFIA data, most prominent at 25\,$\mu$m. It is north of G025.3824-00.1812, and is associated with a near-infrared source seen in the Spitzer-IRAC images. We label this source 1 in Figure~\ref{fig:W42_2}. Its emission cannot be completely separated from the overall extended emission in the SOFIA images, but SED fits to the photometry show that it is also might be a MYSO with a best fit mass $8\,M_{\sun}$ (Table~\ref{tb:sed}). The accuracy of the mass and luminosity parameters derived from the SED fits for all of these sources is likely to be uncertain due to the unknown contribution of the heating of the dust by the revealed O star W42\,\#1. This also goes for the radio-derived masses of these sources, since W42\,\#1 is thought to be responsible for ionizing the entirety of the large-scale \ion{H}{2} emission of W42. 

In summary, the total radio continuum flux from W42 is equal to the ionizing power of the observed O5-O6.5V star of \citet{2000AJ....119.1860B}, and thus it is likely that this one star, W42\,\#1, is responsible for most of the radio continuum emission seen in the region. The overall Lyman continuum photon rate from the entire region is less than that of Orion. Though the central $r\sim1\arcmin$ ($\sim0.6$\,pc) region has several dozen low mass YSOs \citep{2000AJ....119.1860B, 2015ApJ...811...79D}, the MYSO population is small, consisting of a MYSO associated with W42-MME, a UC\ion{H}{2} region G025.3824-00.1812 hosting a young MYSO, a weak radio-emitting MYSO associated with infrared source 1, and potentially another UC\ion{H}{2} region G025.3809-00.1815 hosting another MYSO. As we will discuss more in Section~\ref{sec:reject}, this MYSO population is modest when compared to the a G\ion{H}{2} regions we have previously studied.

\section{Reassessing the G\ion{H}{2} Region Census}\label{sec:reassess}

In light of the fact that the new distances for Sgr\,D and W42 have led to their demotion from G\ion{H}{2} region status, we were motivated to perform in-depth literature searches for recent and more accurate distance measurements toward each of the 56 sources in the \citet{2004MNRAS.355..899C} census. In this section, we will briefly describe the different methods of distance determination, describe the calculation of $N_{LyC}$ performed, and list the sources that do and do not qualify as G\ion{H}{2} regions based upon these updated calculations.     

\subsection{Updating Distance Measurements}\label{sec:dist}

An accurate determination of the distance toward each G\ion{H}{2} region is critically tied to their identity as a G\ion{H}{2} region, since the derived luminosity, and more importantly derived Lyman continuum photon rate, are proportional to the square of their distance from the Sun. From our literature search on each source we compiled the latest and/or most reliable distance estimate for each source. The vast majority of sources have had their accepted distances adjusted since the publications of \citet{2004MNRAS.355..899C}. 

In most cases there are several distance estimates, and therefore we chose to adopt the distance derived via the measurements that were most precise. Though measurements made by any particular methodology will have a range of precisions, in general we can rank the methods of determining distance from most accurate to least accurate as: 1) trigonometric parallax measurements either via circumstellar masers, or via GAIA measurements of the revealed stellar clusters associated with the ionized region, 2) spectrophotometric measurements, 3) kinematically-derived distances outside the solar circle or inside the solar circle but at a tangent point, 4) sources with large differences in their near/far kinematic distance values but for which there is optical or H$\alpha$ emission (an indication of low interstellar extinction) and thus indicating the near distance; 5) an assumed association of the G\ion{H}{2} region with a nearby object/region at similar $v_{lsr}$ and for which a distance is more accurately known (e.g., regions near the Galactic Center which are assumed to be at the same distance as Sgr\,A\text{*} derived via maser parallax measurements), 6) kinematically-derived distances where sources have measured \ion{H}{1} (or other species) absorption line velocity observations that help resolve the kinematic distance ambiguity, and finally, 7) kinematically-derived distances with near/far ambiguity but no supporting deconflicting observations, or having conflicting observations from methods with similar accuracy, which means the kinematic distance ambiguity cannot be clearly resolved. For more detailed explanations of all of these methods, see the thorough discussion on distance measurements and their accuracies in \citet{2018MNRAS.473.1059U}. 

Since there are a variety of Galactic rotation curve models, if our literature search only turned up kinematic distance measurements for a source, we recorded the v$_{lsr}$ values from line measurements (typically radio H$\alpha$ lines) that had the best reported precision. We then derived kinematic distances ourselves so that all sources have kinematic distance determinations  performed in a consistent way. For this we chose to apply the Monte Carlo kinematic distance method of \citet{2018ApJ...856...52W} which utilizes the \citet{2014ApJ...783..130R} rotation curve and updated solar motion parameters (the most important parameter of which to point out is the assumed distance to the Galactic Center of $R_0=8.34\pm0.16$\,kpc). 

Table~\ref{tb:inputs} lists the old distances (column 4) from \citet{2004MNRAS.355..899C} along with the newly adopted distances with their associated errors (column 5), the reference for the new distance measurement and error (column 6), and which of the methodologies listed above was employed in the distance measurement (column 7). For sources with only kinematic distance determinations, Table~\ref{tb:kinematic} lists the adopted v$_{lsr}$ value from the literature, our calculated near/far (or tangent) distances found via the Monte Carlo kinematic distance method, and the method used to achieve a kinematic distance ambiguity resolution (KDAR) to get the adopted distance listed in Table~\ref{tb:inputs}. Appendix~\ref{sec:app} provides an in-depth discussion of the distance measurements towards each source, and the reasoning behind the adopted distances tabulated in Table~\ref{tb:inputs}.

\subsection{Recalculating Lyman Conntinuum Photon Rate}\label{sec:lyman}

Once new distance measurements towards all sources were obtained, we recalculated the Lyman continuum photon rate of each source. We used the relationship for the observed Lyman continuum photon rate, $N^{\prime}_{LyC}$, defined in \citet{1974AA....32..269M}:
\begin{eqnarray}
N^{\prime}_{LyC} = 4.761\times10^{48} a(\nu, T_{e})^{-1} \left(\frac{\nu}{\mathrm{GHz}}\right)^{0.1} \nonumber \\
\times \left(\frac{T_{e}}{\mathrm{K}}\right)^{-0.45} \left(\frac{S_{\nu}}{\mathrm{Jy}}\right) \left(\frac{d}{\mathrm{kpc}}\right)^{2}~\mathrm{photons~s^{-1}},
\label{equ:Npc}
\end{eqnarray} 

\noindent
where $\nu$ is frequency, $T_e$ is electron temperature, $S_{\nu}$ is radio flux density, and $d$ is distance to the source. The $a(\nu, T_{e})$ term is approximately equal to 1, but we directly calculate it from the electron temperature through the expression given by \citet{1967ApJ...147..471M}: 
\begin{eqnarray}
a(\nu, T_{e}) = 0.366 \left( \frac{\nu}{\mathrm{GHz}} \right)^{0.1} \left(\frac{T_{e}}{\mathrm{K}}\right)^{-0.15} \nonumber \\
\times \left\{ \mathrm{ln} \left( 4.995\times10^{-2} \left(\frac{\nu}{\mathrm{GHz}}\right)^{-1} \right) + 1.5~\mathrm{ln} \left(\frac{T_{e}}{\mathrm{K}}\right)  \right\}.
\label{equ:ant}
\end{eqnarray}
\noindent
Where available, we updated the $T_e$ values tabulated by \citet{2004MNRAS.355..899C} with more precise measurements. All new measurements adopted here come from \citet{2019ApJ...887..114W}, which expands upon, and in some cases updates (with higher quality observations) the work of \citet{2015ApJ...806..199B}. These adopted $T_e$ values with their errors are given in column 10 of Table~\ref{tb:inputs}, and the reference for these measurements are given in column 11. 

From here we use the methodology in \citet{1978AA....66...65S} to estimate the fraction of photons lost to dust absorption to correct the observed Lyman continuum photon rate, $N^{\prime}_{LyC}$, and derive the intrinsic Lyman continuum photon rate, $N_{LyC}$. This calculation requires first calculating the galactocentric distance of each source, which we calculated with the assumption that the Sun is 8.34$\pm$0.16\,kpc from the Galactic Center using the values from \citet{2014ApJ...783..130R} which is the value also used in the calculation adopted for obtaining kinematic distances; this is also different from the work of \citet{2004MNRAS.355..899C} who used a distance of $R_0=8.0$\,kpc. We simply use the law of cosines to determine the galactocentric distance to each source:
\begin{eqnarray}
R_{GC} = [ d^2 + R_{0}^2 - 2\,d\,R_{0}\,cos(l) ]^{0.5}~\mathrm{kpc},
\label{equ:RGC}
\end{eqnarray}

\noindent
where $l$ is the source Galactic longitude. Next, one calculates the helium absorption cross-section parameter, $x\sigma_{He}/x_1\sigma_{\nu}$, defined and discussed in \citet{1978AA....66...65S}, which was found to have the following empirically-derived relationship with galactocentric distance \citep[see also][]{1978AA....70..719C}:
\begin{eqnarray}
x\sigma_{He}/x_1\sigma_{\nu} = 
\begin{dcases}
   5.0-0.4\,R_{GC}\,(\pm2.0), & \text{for } R_{GC} < 10\,\mathrm{kpc}\\
   1.0\,(\pm1.0),              & \text{for } R_{GC} > 10\,\mathrm{kpc}.
\end{dcases}
\label{equ:xs}
\end{eqnarray}

\noindent
Calculation of intrinsic Lyman continuum photon rate further requires the calculation of the electron density of the source, $N_e$, which we took from \citet{1969ApJ...156..269S}:
\begin{eqnarray}
N_e = 98.152~a(\nu, T_{e})^{-0.5} \left(\frac{\nu}{\mathrm{GHz}}\right)^{0.05} \left(\frac{T_{e}}{\mathrm{K}}\right)^{0.175} \nonumber \\
\times  \left(\frac{S_{\nu}}{\mathrm{Jy}}\right)^{0.5} \left(\frac{d}{\mathrm{kpc}}\right)^{-0.5}\left(\frac{\theta}{\mathrm{arcmin}}\right)^{-1.5}~\mathrm{cm^{-3}},
\label{equ:Ne}
\end{eqnarray} 

\noindent
where $\theta$ is the half-power width of a Gaussian fit to the radio source.  

Using the resultant values derived for $N^{\prime}_{LyC}$ from Equation \ref{equ:Npc}, $N_e$ from Equation \ref{equ:Ne}, and $x\sigma_{He}/x_1\sigma_{\nu}$ from Equation \ref{equ:xs}, we plug those values into the equation for H-photon absorption optical depth given by \citet{1978AA....66...65S}:
\begin{eqnarray}
\tau_H = 3.4\times10^{-18}\,(N^{\prime}_{LyC}\,N_e)^{1/3}\,a_o^{-1}\,(x\sigma_{He}/x_1\sigma_{\nu}),  
\label{equ:th}
\end{eqnarray}

\noindent
where $a_o$ is the ratio of the absorption cross sections for He-photons and H-photons. We use the assumption from \citet{1978AA....66...65S} and assume a value to 6.0$\pm$1.0.

\citet{1978AA....66...65S} give a table of values for the fraction of Lyman continuum photons absorbed by the gas, $f_{net}$, and their corresponding values of $\tau_H$ (as well as other parameters). We fit these data in the table with a functional form, given by the equation: 
\begin{eqnarray}
f_{net} = -0.234\,\mathrm{ln}(\tau_H) + 0.259.  
\label{equ:f}
\end{eqnarray}

\noindent
This equation reproduces the $f_{net}$ values for the G\ion{H}{2} regions tabulated in \citet{1978AA....66...65S} to within 6\% (using only the parameters from that work). Finally, we calculate the intrinsic Lyman continuum photon rate using:
\begin{eqnarray}
N_{LyC} = N^{\prime}_{LyC}/f_{net}.  
\label{equ:final}
\end{eqnarray}

\begin{deluxetable*}{lrrcccccccc}
\tabletypesize{\tiny}
\tablecolumns{11}
\tablewidth{0pt}
\tablecaption{Updated List of Assumed G\ion{H}{2} Region Properties}
\tablehead{\colhead{(1)} &
	   \colhead{(2) } &
	   \colhead{(3)} &
	   \colhead{(4)} &
	  \colhead{(5)} &
	   \colhead{(6)} &
	   \colhead{(7) } &
	   \colhead{ (8) } &
	   \colhead{(9) } &
	   \colhead{(10) } &
	   \colhead{ (11)}\\
\colhead{Name}&
	   \colhead{ $\ell$ } &
	   \colhead{ $b$} &
	   \colhead{ Old $d$ } &
	  \colhead{ New $d$ } &
	   \colhead{ Ref. } &
	   \colhead{ Meth. } &
	   \colhead{ $S_{6cm}$ } &
	   \colhead{ $\theta_{6cm}$ } &
	   \colhead{ $T_e$} &
	   \colhead{ Ref. } \\
\colhead{}&
	   \colhead{ ($\arcdeg$) } &
	   \colhead{ ($\arcdeg$)} &
	   \colhead{ (kpc) } &
	  \colhead{ (kpc) } &
	   \colhead{ } &
	   \colhead{ } &
	   \colhead{ (Jy)} &
	   \colhead{ ($\arcmin$) } &
	   \colhead{ } &
	   \colhead{ (K) } 
}
\startdata
G0.361-0.780	&0.361		&-0.780		&8.0	&8.23$^{+0.20}_{-0.17}$		&DWB80	&Kin	&7.4$\pm$0.7		&5.2$\pm$0.4	&5100$\pm$2500			&DWB80\\
G0.394-0.540	&0.394		&-0.540		&8.0	&8.23$^{+0.18}_{-0.15}$		&DWB80	&Kin	&8.8$\pm$0.9		&5.6$\pm$0.4	&7500$\pm$2500			&DWB80 \\
G0.489-0.668	&0.489		&-0.668		&8.0	&8.25$^{+0.25}_{-0.33}$		&DWB80	&Kin	&8.4$\pm$0.8		&3.7$\pm$0.3	&5500$\pm$2500			&DWB80 \\
Sgr\,B1				&0.518		&-0.065		&8.0	&7.80$^{+0.80}_{-0.70}$		&RMZ09	&Sgr\,B2*	&35.1$\pm$3.5		&3.7$\pm$0.3	&7200$\pm$2000		&DWB80 \\
G0.572-0.628	&0.572		&-0.628		&8		&8.18$^{+0.25}_{-0.18}$		&DWB80	&Kin	&7.2$\pm$0.7		&2.5$\pm$0.2	&6200$\pm$2500			&DWB80 \\
Sgr\,D				&1.149		&-0.062		&8.0	&2.36$^{+0.58}_{-0.39}$		&SON17	&Mpara	&19.3$\pm$1.9		&5.5$\pm$0.4	&5000$\pm$500			&WWB83 \\
G2.303+0.243	&2.303		&0.243		&14.3	&13.48$^{+0.98}_{-0.76}$	&L89	&	Kin	&7.3$\pm$0.7		&5.9$\pm$0.5	&3700$\pm$2500			&DWB80 \\
G3.270-0.101	&3.270		&-0.101		&14.3	&14.33$^{+0.76}_{-0.82}$	&JDD13	&Kin	&9.9$\pm$1.0		&5.7$\pm$0.4	&7440$\pm$280			&WBA19 \\
G4.412+0.118	&4.412		&0.118		&14.6	&14.97$^{+0.77}_{-0.58}$	&RWJ16	&Kin	&10.4$\pm$1.0		&4.9$\pm$0.4	&5700$\pm$2500			&DWB80 \\
M8					&5.973		&-1.178		&2.8	&1.34$\pm$0.07		 		&RPB20	&Gpara	&113.4$\pm$11.3		&7.5$\pm$0.6	&8180$\pm$70			&WBA19 \\
G8.137+0.228	&8.137		&0.228		&13.5	&3.38$^{+0.28}_{-0.36}$	&QRB06	&Kin	&8.2$\pm$0.8		&1.8$\pm$0.1	&7090$\pm$60			&WBA19 \\
W31-South			&10.159		&-0.349		&4.5	&3.40$\pm$0.3				&RDC01	&Spec	&66.3$\pm$6.6		&2.9$\pm$0.2	&6830$\pm$30			&WBA19 \\
W31-North			&10.315		&-0.150		&15.0	&1.75$\pm$0.25				&DZS15	&Spec	&20.5$\pm$2.1		&3.1$\pm$0.2	&6800$\pm$40			&WBA19 \\
M17					&15.032		&-0.687		&2.4	&1.98$^{+0.14}_{-0.12}$		&XMR11	&Mpara	&844.5$\pm$84.5		&4.5$\pm$0.3	&9280$\pm$120			&WBA19 \\
G20.733-0.087	&20.733		&-0.087		&11.8	&11.69$^{+0.34}_{-0.44}$	&QRB06	&Kin	&19.5$\pm$2.0		&6.1$\pm$0.5	&5590$\pm$90			&WBA19 \\
W42					&25.382		&-0.177		&11.5	&2.20$^{+0.80}_{-0.60}$		&BCD00	&Spec	&29.5$\pm$3.0		&3.7$\pm$0.3	&7460$\pm$70			&WBA19 \\
G29.944-0.042	&29.944		&-0.042		&6.2	&5.71$^{+0.50}_{-0.42}$		&ZMS14	&Mpara	&25.5$\pm$2.6		&3.7$\pm$0.3	&6510$\pm$90			&WBA19 \\
W43					&30.776		&-0.029		&6.2	&5.49$^{+0.39}_{-0.34}$		&ZMS14	&Mpara	&62.2$\pm$6.2		&4.1$\pm$0.3	&6567$\pm$30			&WBA19 \\
G32.80+0.19			&32.797		&0.192		&12.9	&12.85$^{+0.44}_{-0.34}$	&QRB06	&Kin	&5.8$\pm$0.6		&2.3$\pm$0.2	&8625$\pm$49			&WBA19 \\
W49A				&43.169		&0.002		&11.8	&11.11$^{+0.79}_{-0.69}$	&ZRM13	&Mpara	&69.0$\pm$6.9		&3.0$\pm$0.2	&7876$\pm$35				&WBA19 \\
G48.596+0.042	&48.596		&0.042		&9.8	&10.75$^{+0.61}_{-0.55}$	&ZRM13	&Mpara	&12.2$\pm$1.5		&4.2$\pm$0.3	&7800$\pm$2500			&DWB80 \\
G48.9-0.3			&48.930		&-0.286		&5.5	&5.62$^{+0.59}_{-0.49}$		&NKO15	&Mpara	&24.3$\pm$2.9		&4.4$\pm$0.3	&8440$\pm$60			&WBA19 \\
W51A:G49.4-0.3		&49.384		&-0.298		&5.5	&5.41$^{+0.31}_{-0.28}$		&SRB10	&Mpara	&27.2$\pm$3.3		&2.8$\pm$0.2	&8585$\pm$65			&WBA19 \\
W51A:G49.5-0.4		&49.486		&-0.381		&5.5	&5.41$^{+0.31}_{-0.28}$		&SRB10	&Mpara	&110.4$\pm$13.2		&2.8$\pm$0.2	&7166$\pm$25			&WBA19 \\
K3-50\,(W58A)		&70.300		&1.600		&8.6	&7.64$^{+0.81}_{-0.54}$		&QRB06	&KOG	&13.0$\pm$1.6		&2.1$\pm$0.2	&10810$\pm$130			&WBA19 \\
DR7					&79.293		&1.296		&8.3	&7.30$^{+0.84}_{-0.72}$		&QRB06	&KOG	&15.8$\pm$1.9		&3.4$\pm$0.3	&8693$\pm$86			&WBA19 \\
W3					&133.720	&1.210		&4.2	&2.30$^{+0.19}_{-0.16}$		&NGD19	&Gpara	&74.7$\pm$9.0		&3.4$\pm$0.3	&8977$\pm$38			&WBA19 \\
RCW42				&274.013	&-1.141		&6.4	&5.97$^{+0.90}_{-0.72}$		&CH87	&KOG	&39.9$\pm$4.0		&2.9$\pm$0.2	&7900$^{+400}_{-400}$		&CH87  \\
RCW46           	&282.023	&-1.180		&5.9	&5.77$^{+0.77}_{-0.77}$		&CH87	&KOG	&40.9$\pm$4.1		&3.8$\pm$0.3	&6200$^{+300}_{-400}$		&CH87  \\
RCW49				&284.301	&-0.344		&4.7	&4.16$\pm$0.27		 		&VKB13	&Spec	&263.2$\pm$26.3		&7.4$\pm$0.6	&8000$^{+300}_{-300}$		&CH87  \\
NGC3372				&287.379	&-0.629		&2.5	&2.3$\pm$0.1 		 		&S06	&$\eta$\,Car*	&145.6$\pm$14.6		&7.0$\pm$0.5	&7200$^{+400}_{-500}$	&CH87  \\
G289.066-0.357	&289.066	&-0.357		&7.9	&7.15$^{+0.54}_{-0.93}$		&CH87	&KOG	&16.4$\pm$1.6		&6.0$\pm$0.5	&8500$^{+600}_{-700}$		&CH87  \\
NGC3576				&291.284	&-0.712		&3.1	&2.77$\pm$0.31		 		&BP18	&Gpara	&113.0$\pm$11.3		&2.5$\pm$0.2	&7500$^{+400}_{-400}$		&CH87  \\
NGC3603				&291.610	&-0.528		&7.9	&7.20$\pm$0.10 		 		&DMW19	&Gpara	&261.0$\pm$26.1		&6.9$\pm$0.5	&6900$^{+100}_{-100}$		&CH87  \\
G298.227-0.340	&298.227	&-0.340		&10.4	&12.4$\pm$1.7 		 		&DNB16	&Spec	&47.4$\pm$4.7		&3.8$\pm$0.3	&8600$^{+600}_{-700}$		&CH87  \\
G298.862-0.438	&298.862	&-0.438		&10.4	&12.4$\pm$1.7				&DNB16	&Spec	&42.4$\pm$4.2		&3.8$\pm$0.3	&6600$^{+100}_{-100}$		&CH87  \\
G305.359+0.194	&305.359	&0.194		&3.5	&3.59$\pm$0.85				&BP18	&Gpara	&56.4$\pm$5.6		&3.5$\pm$0.3	&5100$^{+200}_{-300}$		&CH87  \\
G319.158-0.398	&319.158	&-0.398		&11.5	&11.26$^{+0.35}_{-0.42}$	&CH87	&Kin	&11.2$\pm$1.1		&5.9$\pm$0.5	&6300$^{+400}_{-400}$		&CH87  \\
G319.392-0.009	&319.392	&-0.009		&11.5	&11.78$^{+0.34}_{-0.42}$	&CH87	&Kin	&8.9$\pm$0.9		&3.5$\pm$0.3	&7700$^{+1000}_{-1300}$	&CH87  \\
G320.327-0.184	&320.327	&-0.184		&12.6	&0.64$^{+0.38}_{-0.27}$		&CH87	&Kin	&6.3$\pm$0.6		&5.0$\pm$0.4	&5700$^{+400}_{-400}$		&CH87  \\
RCW97				&327.304	&-0.552		&3.0	&2.98$^{+0.23}_{-0.36}$		&WMS06	&Kin	&64.9$\pm$6.5		&2.9$\pm$0.2	&4700$^{+300}_{-300}$		&CH87  \\
G327.993-0.100	&327.993	&-0.100		&11.4	&2.80$^{+0.31}_{-0.31}$		&CH87	&Kin	&5.9$\pm$0.6		&2.7$\pm$0.2	&6000$^{+600}_{-700}$		&CH87  \\
G330.868-0.365	&330.868	&-0.365		&10.8	&3.44$^{+0.47}_{-0.36}$		&CH87	&Kin	&14.7$\pm$1.5		&4.0$\pm$0.3	&4900$^{+400}_{-500}$		&CH87  \\
G331.324-0.348	&331.324	&-0.348		&10.8	&3.29$\pm$0.58		 		&PAC12	&Spec	&6.5$\pm$0.7		&1.8$\pm$0.1	&3500$^{+200}_{-200}$		&CH87  \\
G331.354+1.072	&331.354	&1.072		&10.8	&4.50$^{+0.55}_{-0.34}$		&CH87	&Kin	&6.7$\pm$0.7		&4.9$\pm$0.4	&5400$^{+400}_{-400}$		&CH87  \\
G331.529-0.084	&331.529	&-0.084		&10.8	&7.31$\pm$2.19				&CH87	&Kin	&47.1$\pm$4.7		&4.0$\pm$0.3	&6200$^{+400}_{-400}$		&CH87  \\
G333.122-0.446	&333.122	&-0.446		&3.5	&2.60$\pm$0.20 				&FBD05	&Spec	&49.5$\pm$5.0		&4.5$\pm$0.3	&5800$^{+500}_{-600}$		&CH87  \\
G333.293-0.382	&333.293	&-0.382		&3.5	&2.60$\pm$0.70 				&RAO09	&Spec	&45.5$\pm$4.6		&3.8$\pm$0.3	&6300$^{+300}_{-400}$		&CH87  \\
G333.610-0.217	&333.610	&-0.217		&3.1	&2.54$\pm$0.71				&RPB20	&Gpara	&116.2$\pm$11.6		&3.7$\pm$0.3	&6200$^{+200}_{-300}$		&CH87  \\
G338.398+0.164	&338.398	&0.164		&13.1	&13.29$^{+0.25}_{-0.45}$	&CH87	&Kin	&25.7$\pm$2.6		&5.5$\pm$0.4	&6600$^{+400}_{-400}$		&CH87  \\
G338.400-0.201	&338.400	&-0.201		&15.7	&15.71$^{+0.58}_{-0.40}$	&CH87	&KOG	&6.3$\pm$0.6		&3.3$\pm$0.3	&9100$^{+700}_{-900}$		&CH87  \\
G345.555-0.043	&345.555	&-0.043		&15.2	&15.28$^{+0.57}_{-0.35}$	&CH87	&Kin	&15.1$\pm$1.5		&5.5$\pm$0.4	&6500$^{+400}_{-400}$		&CH87  \\
G345.645+0.009	&345.645	&0.009		&15.2	&14.97$^{+0.39}_{-0.45}$	&CH87	&Kin	&11.2$\pm$1.1		&4.2$\pm$0.3	&7800$^{+500}_{-500}$		&CH87  \\
G347.611+0.204	&347.611	&0.204		&6.6	&7.90$\pm$0.8				&BGH12	&Spec	&23.2$\pm$2.3		&6.1$\pm$0.5	&4000$^{+300}_{-400}$		&CH87  \\
G351.467-0.462	&351.467	&-0.462		&13.7	&3.24$^{+0.34}_{-0.26}$		&CH87	&Spec*	&4.7$\pm$0.5		&2.7$\pm$0.2	&7460$\pm$120			&WBA19 \\
Sgr C				&359.429	&-0.090		&8.0	&8.34$^{+0.15}_{-0.17}$		&CH87	&Tang	&19.3$\pm$1.9		&4.3$\pm$0.3	&9300$^{+500}_{-500}$		&CH87
\enddata
\tablecomments{References in Column 6 are for distance measurements, and Column 11 are for the electron temperature measurements, and use the following abbreviations: BCD00 \citep{2000AJ....119.1860B},
BDC01 \citep{2001AJ....121.3149B},
BGH12 \citep{2012AA...546A.110B}, 
BP18 \citep{2018ApJ...864..136B}, 
CH87 \citep{1987AA...171..261C},
DMW19 \citep{2019MNRAS.486.1034D}, 
DNB16 \citep{2016AA...589A..69D},
DWB80 \citep{1980AAS...40..379D}, 
DZS15 \citep{2015AA...582A...1D}, 
FBD05 \citep{2005AJ....129.1523F},
JDD13 \citep{2013ApJ...774..117J},
L89 \citep{1989ApJS...71..469L},
NGD19 \citep{2019MNRAS.487.2771N},
NKO15 \citep{2015PASJ...67...65N},
PAC12 \citep{2012MNRAS.423.2425P}, 
QRB06 \citep{2006ApJ...653.1226Q},
RAO09 \citep{2009MNRAS.394..467R},
RMZ09 \citep{2009ApJ...705.1548R}, 
RPB20 \citep{2020AA...633A.155R},
RWJ16 \citep{2016PASA...33...30R},
S06 \citep{2006MNRAS.367..763S},  
SON17 \citep{2017PASJ...69...64S},
SRB10 \citep{2010ApJ...720.1055S},
VKB13 \citep{2013AJ....145..125V},
WBA19 \citep{2019ApJ...887..114W},
WMS06 \citep{2006AA...454L..91W},
WWB83 \citep{1983AA...127..211W},
XMR11 \citep{2011ApJ...733...25X},  
ZMS14 \citep{2014ApJ...781...89Z}, and
ZRM13 \citep{2013ApJ...775...79Z}.
Column 7: The methods used for determining distances are:  `Mpara/Gpara' - maser/GAIA parallax; `Spec' - Spectrophotometric/Photometric; `Kin' - Kinematic distance with a resolved distance ambiguity (see Table~\ref{tb:kinematic} for more details); `KOG/Tang' - Kinematic distance with no ambiguity because it resides in the outer Galaxy or at a tangent point;  `Sgr\,B2' - assumed to be at the maser parallax-derived distance of Sgr\,B2, `$\eta$ Car' - assumed to be at the distance of $\eta$ Car.}
\tablenotetext{*}{Distance derived spectrophotometrically by \citet{2006AA...455..923B}, but not quoted a formal uncertainty. The spectrophotometric value is the same as the kinematically-derived value, so we quote the kinematic distance with it's formal error.}
\label{tb:inputs}
\end{deluxetable*}

\begin{deluxetable*}{lccccccc}
\tabletypesize{\scriptsize}
\tablewidth{0pt}
\tablecaption{Parameters Used in Determining Distances to Sources That Only Have Kinematic Information}
\tablehead{\colhead{Name} &
	   \colhead{ $v_{lsr}$ } &
	  \colhead{ $v_{lsr}$ } &
	   \colhead{ Near } &
	   \colhead{ Far } &
	   \colhead{ KDAR } &
	   \colhead{ KDAR  } &
	   \colhead{ KDAR } \\
       \colhead{   } &
	   \colhead{ (km/s) } &
	   \colhead{ Reference} &
	   \colhead{ Distance } &
	  \colhead{ Distance } &
	   \colhead{ } &
	   \colhead{ Method } &
	   \colhead{ Reference } 
}
\startdata
G0.361-0.780	&	20.0$\pm$5.0	&	Downes et al. (1980)	&	$8.23^{+0.20}_{-0.17}$	&	$8.45^{+0.15}_{-0.25}$	&	Near	&	H$\alpha$	&	Russeil (2003)	\\
G0.394-0.540	&	24.0$\pm$5.0	&	Downes et al. (1980)	&	$8.23^{+0.18}_{-0.15}$	&	$8.30^{+0.23}_{-0.12}$	&	Near	&	H$\alpha$	&	Russeil (2003)	\\
G0.489-0.668	&	17.0$\pm$5.0	&	Downes et al. (1980)	&	$8.25^{+0.25}_{-0.33}$	&	$8.44^{+0.33}_{-0.17}$	&	Near	&	H$\alpha$	&	Russeil (2003)	\\
G0.572-0.628	&	20.0$\pm$5.0	&	Downes et al. (1980)	&	$8.18^{+0.25}_{-0.18}$	&	$8.48^{+0.20}_{-0.20}$	&	Near	&	H$\alpha$	&	Russeil (2003)	\\
G2.303+0.243	&	4.9$\pm$0.7	&	Lockman (1989)	&	$3.14^{+0.74}_{-0.89}$	&	$13.48^{+0.98}_{-0.76}$	&	Far	&	OH	&	Russeil (2003)	\\
G3.270-0.101	&	4.9$\pm$0.8	&	Caswell \& Haynes (1987)	&	$2.45^{+0.56}_{-0.81}$	&	$14.33^{+0.76}_{-0.82}$	&	Far	&	\ion{H}{1}	&	Jones et al. (2013)	\\
G4.412+0.118	&	4.23$\pm$0.25	&	Rathborne et al. (2016)	&	$1.79^{+0.43}_{-0.75}$	&	$14.97^{+0.77}_{-0.58}$	&	Far	&	\ion{H}{1}	&	Jones et al. (2013)	\\
G8.137+0.228	&   20.31$\pm$0.06		& Quireza et al. (2006)	&3.38$^{+0.28}_{-0.36}$	   &13.17$^{+0.44}_{-0.39}$	& Near	& \ion{H}{1}		& Jones et al. (2013) \\
G20.733-0.087	&	55.96$\pm$0.04	&	Quireza et al. (2006)	&	$3.85^{+0.39}_{-0.27}$	&	$11.69^{+0.34}_{-0.44}$	&	Far	&	\ion{H}{1}	&	Urquhart et al. (2018)	\\
G32.80+0.19	&	15.46$\pm$0.15	&	Quireza et al. (2006)	&	$1.10^{+0.25}_{-0.33}$	&	$12.85^{+0.44}_{-0.34}$	&	Far	&	Masers$^{a}$	&	Zhang et al. (2019)	\\
G319.158-0.398	&	21.0$\pm$1.0	&	Caswell \& Haynes (1987)	&	$1.41^{+0.21}_{-0.40}$	&	$11.26^{+0.35}_{-0.42}$	&	Far	&	\ion{H}{1}	&	Urquhart et al. (2012)	\\
G319.392-0.009	&	-14.0$\pm$1.0	&	Caswell \& Haynes (1987)	&	$1.01^{+0.23}_{-0.43}$	&	$11.78^{+0.34}_{-0.42}$	&	Far	&	\ion{H}{1}	&	Urquhart et al. (2012)	\\
G320.327-0.184	&	-11.0$\pm$1.0	&	Caswell \& Haynes (1987)	&	$0.64^{+0.38}_{-0.27}$	&	$12.02^{+0.54}_{-0.30}$	&	Near?$^{b}$	&	\ion{H}{1}$^{b}$	&	Urquhart et al. (2018)	\\
RCW97	&	-47.5$\pm$0.1	&	Wyrowski et al. (2006)	&	$2.98^{+0.23}_{-0.36}$	&	$11.14^{+0.34}_{-0.41}$	&	Near	&	\ion{H}{1}	&	Urquhart et al. (2012)	\\
G327.993-0.100	&	-45.0$\pm$1.0	&	Caswell \& Haynes (1987)	&	$2.80^{+0.31}_{-0.31}$	&	$11.28^{+0.38}_{-0.36}$	&	Near	&	\ion{H}{1}	&	Urquhart et al. (2018)	\\
G330.868-0.365	&	-56.0$\pm$1.0	&	Caswell \& Haynes (1987)	&	$3.44^{+0.47}_{-0.36}$	&	$11.02^{+0.47}_{-0.36}$	&	Near	&	Optical	&	Paladini et al. (2004)	\\
G331.354+1.072	&	-79.0$\pm$1.0	&	Caswell \& Haynes (1987)	&	$4.50^{+0.55}_{-0.34}$	&	$9.98^{+0.46}_{-0.46}$	&	Near	&	\ion{H}{1}	&	Urquhart et al. (2012)	\\
G331.529-0.084	&	-89.0$\pm$1.0	&	Caswell \& Haynes (1987)	&	$7.31\pm2.19^{c}$	&	\nodata	&	Tangent?$^{c}$	&	$^{c}$	&	Merello et al. (2013)	\\
G338.398+0.164	&	-29.0$\pm$1.0	&	Caswell \& Haynes (1987)	&	$2.32^{+0.20}_{-0.31}$	&	$13.29^{+0.25}_{-0.45}$	&	Far?$^{b}$	&	\ion{H}{1}$^{b}$	&	Urquhart et al. (2018)	\\
G345.555-0.043	&	-6.0$\pm$1.0	&	Caswell \& Haynes (1987)	&	$0.60^{+0.44}_{-0.19}$	&	$15.28^{+0.57}_{-0.35}$	&	Far	&	CO/H2CO	&	Caswell \& Haynes (1987)	\\
G345.645+0.009	&	-10.0$\pm$1.0	&	Caswell \& Haynes (1987)	&	$1.10^{+0.41}_{-0.19}$	&	$14.97^{+0.39}_{-0.45}$	&	Far	&	\ion{H}{1}	&	Urquhart et al. (2012)	
\enddata
\tablecomments{`KDAR' stands for kinematic distance ambiguity resolution. `H$\alpha$' - the presence of H$\alpha$ emission indicating the near kinematic  distance; `OH', `\ion{H}{1}', and `CO/H$_2$CO' - the presence of these absorption features indicating near or far distance, and `Optical' - the presence of optical emission from the region indicates near distance. }
\tablenotetext{a}{Low precision maser parallax measurements consistent with the higher precision kinematic far distance.}
\tablenotetext{b}{Conflicting absorption line measurements pointing to both near and far distances.}
\tablenotetext{c}{Conflicting information points to near, far, and tangent distance, therefore tangent is quoted with error that covers both near and far distance.}
\label{tb:kinematic}
\end{deluxetable*}

\begin{deluxetable*}{lccccc}
\tabletypesize{\scriptsize}
\tablecolumns{6}
\tablewidth{0pt}
\tablecaption{Derived G\ion{H}{2} and \ion{H}{2} Region Properties}\label{tb:outputs}
\tablehead{
\colhead{Name}&
	   \colhead{ $N^{\prime}_{LyC}$} &
	  \colhead{ $N_e$ } &
	   \colhead{ $R_{GC}$ } &
	   \colhead{ $N_{LyC}$} &
	   \colhead{ Is G\ion{H}{2}? } \\
\colhead{}&
	   \colhead{ log(s$^{-1}$) } &
	  \colhead{ (cm$^{-3}$) } &
	   \colhead{ (kpc) } &
	   \colhead{ log(s$^{-1}$) } &
	   	   \colhead{ }
}
\startdata
G0.361-0.780	&49.76$^{+0.10}_{-0.11}$   &  38  $^{+  5}_{-  6}$  & 0.06$^{+0.23}_{-0.01}$ &50.06$^{+0.17}_{-0.15}$	&Likely	  \\
G0.394-0.540	&49.76$^{+0.09}_{-0.07}$   &  38  $^{+  6}_{-  4}$  & 0.06$^{+0.21}_{-0.01}$ &50.07$^{+0.15}_{-0.12}$	&Likely	  \\
G0.489-0.668	&49.78$^{+0.12}_{-0.09}$   &  65  $^{+ 13}_{-  8}$  & 0.08$^{+0.32}_{-0.01}$ &50.15$^{+0.17}_{-0.16}$	&Likely	  \\
Sgr\,B1			&50.31$^{+0.14}_{-0.07}$   & 146  $^{+ 22}_{- 20}$  & 0.10$^{+0.83}_{-0.01}$ &50.87$^{+0.22}_{-0.22}$	&Yes      \\
G0.572-0.628	&49.70$^{+0.10}_{-0.09}$   & 117  $^{+ 15}_{- 19}$  & 0.09$^{+0.24}_{-0.00}$ &50.08$^{+0.18}_{-0.15}$	&Likely   \\
Sgr\,D			&49.15$^{+0.19}_{-0.16}$   &  99  $^{+ 18}_{- 13}$  & 5.95$^{+0.45}_{-0.58}$ &49.37$^{+0.19}_{-0.27}$	&No       \\
G2.303+0.243	&50.22$^{+0.15}_{-0.13}$   &  22  $^{+  4}_{-  3}$  & 5.07$^{+1.06}_{-0.67}$ &50.48$^{+0.20}_{-0.19}$	&Yes      \\
G3.270-0.101	&50.30$^{+0.07}_{-0.06}$   &  30  $^{+  4}_{-  3}$  & 6.07$^{+0.73}_{-0.88}$ &50.55$^{+0.16}_{-0.13}$	&Yes      \\
G4.412+0.118	&50.40$^{+0.12}_{-0.09}$   &  36  $^{+  5}_{-  5}$  & 6.78$^{+0.70}_{-0.67}$ &50.69$^{+0.19}_{-0.18}$	&Yes      \\
M8				&49.29$^{+0.07}_{-0.06}$   & 227  $^{+ 31}_{- 28}$  & 7.00$^{+0.18}_{-0.17}$ &49.52$^{+0.14}_{-0.15}$	&No       \\
G8.137+0.228	&48.99$^{+0.07}_{-0.12}$   & 318  $^{+ 42}_{- 27}$  & 5.04$^{+0.37}_{-0.33}$ &49.20$^{+0.15}_{-0.17}$	&No       \\
W31-South		&49.90$^{+0.08}_{-0.09}$   & 446  $^{+ 53}_{- 56}$  & 5.01$^{+0.35}_{-0.31}$ &50.32$^{+0.20}_{-0.22}$	&Yes      \\
W31-North		&48.82$^{+0.13}_{-0.14}$   & 303  $^{+ 48}_{- 33}$  & 6.58$^{+0.35}_{-0.24}$ &48.97$^{+0.22}_{-0.16}$	&No       \\
M17				&50.49$^{+0.07}_{-0.07}$   &1114  $^{+144}_{-119}$  & 6.43$^{+0.21}_{-0.19}$ &51.02$^{+0.36}_{-0.25}$	&Yes      \\
G20.733-0.087	&50.48$^{+0.05}_{-0.06}$   &  40  $^{+  6}_{-  4}$  & 4.85$^{+0.30}_{-0.33}$ &50.80$^{+0.16}_{-0.15}$	&Yes      \\
W42				&49.24$^{+0.25}_{-0.29}$   & 237  $^{+ 63}_{- 38}$  & 6.56$^{+0.37}_{-0.83}$ &49.44$^{+0.33}_{-0.35}$	&No       \\
G29.944-0.042	&49.95$^{+0.09}_{-0.08}$   & 142  $^{+ 21}_{- 18}$  & 4.37$^{+0.21}_{-0.17}$ &50.31$^{+0.16}_{-0.19}$	&Yes      \\
W43				&50.30$^{+0.08}_{-0.07}$   & 199  $^{+ 24}_{- 25}$  & 4.53$^{+0.21}_{-0.15}$ &50.76$^{+0.18}_{-0.22}$	&Yes      \\
G32.80+0.19		&49.96$^{+0.05}_{-0.05}$   &  96  $^{+ 16}_{- 11}$  & 7.41$^{+0.31}_{-0.31}$ &50.23$^{+0.13}_{-0.17}$	&Yes      \\
W49A			&50.92$^{+0.08}_{-0.07}$   & 240  $^{+ 31}_{- 26}$  & 7.57$^{+0.53}_{-0.44}$ &51.42$^{+0.25}_{-0.28}$	&Yes      \\
G48.596+0.042	&50.15$^{+0.09}_{-0.10}$   &  62  $^{+  9}_{-  8}$  & 8.18$^{+0.38}_{-0.37}$ &50.38$^{+0.18}_{-0.18}$	&Yes      \\
G48.9-0.3		&49.85$^{+0.11}_{-0.08}$   & 114  $^{+ 15}_{- 14}$  & 6.31$^{+0.12}_{-0.13}$ &50.16$^{+0.15}_{-0.20}$	&Likely   \\
W51A:G49.4-0.3	&49.87$^{+0.08}_{-0.06}$   & 249  $^{+ 27}_{- 33}$  & 6.32$^{+0.14}_{-0.11}$ &50.22$^{+0.15}_{-0.20}$	&Yes      \\
W51A:G49.5-0.4	&50.51$^{+0.08}_{-0.07}$   & 468  $^{+ 72}_{- 48}$  & 6.34$^{+0.13}_{-0.11}$ &51.03$^{+0.24}_{-0.27}$	&Yes      \\
K3-50\,(W58A)	&49.84$^{+0.08}_{-0.11}$   & 220  $^{+ 41}_{- 32}$  & 9.22$^{+0.44}_{-0.31}$ &50.07$^{+0.19}_{-0.17}$	&Likely   \\
DR7				&49.90$^{+0.12}_{-0.10}$   & 117  $^{+ 20}_{- 16}$  &10.10$^{+0.40}_{-0.52}$ &50.10$^{+0.14}_{-0.18}$	&Likely   \\
W3				&49.57$^{+0.09}_{-0.08}$   & 461  $^{+ 76}_{- 62}$  &10.09$^{+0.21}_{-0.22}$ &49.78$^{+0.14}_{-0.17}$	&No       \\
RCW42			&50.17$^{+0.12}_{-0.13}$   & 261  $^{+ 35}_{- 33}$  & 9.95$^{+0.47}_{-0.45}$ &50.42$^{+0.19}_{-0.18}$	&Yes      \\
RCW46           &50.17$^{+0.12}_{-0.12}$   & 170  $^{+ 30}_{- 21}$  & 9.09$^{+0.36}_{-0.36}$ &50.44$^{+0.22}_{-0.20}$	&Yes      \\
RCW49			&50.64$^{+0.07}_{-0.07}$   & 194  $^{+ 33}_{- 21}$  & 8.39$^{+0.12}_{-0.19}$ &51.08$^{+0.18}_{-0.26}$	&Yes      \\
NGC3372			&49.90$^{+0.06}_{-0.06}$   & 216  $^{+ 24}_{- 28}$  & 7.97$^{+0.15}_{-0.16}$ &50.20$^{+0.15}_{-0.19}$	&Yes      \\
G289.066-0.357	&49.87$^{+0.11}_{-0.10}$   &  53  $^{+  8}_{-  7}$  & 8.93$^{+0.36}_{-0.36}$ &50.07$^{+0.14}_{-0.20}$	&Likely   \\
NGC3576			&49.95$^{+0.10}_{-0.11}$   & 786  $^{+140}_{- 88}$  & 7.77$^{+0.16}_{-0.14}$ &50.30$^{+0.26}_{-0.21}$	&Yes      \\
NGC3603			&51.15$^{+0.05}_{-0.05}$   & 159  $^{+ 25}_{- 14}$  & 8.77$^{+0.13}_{-0.09}$ &51.61$^{+0.26}_{-0.25}$	&Yes      \\
G298.227-0.340	&50.84$^{+0.13}_{-0.12}$   & 130  $^{+ 25}_{- 14}$  &11.19$^{+1.30}_{-1.25}$ &51.11$^{+0.20}_{-0.19}$	&Yes      \\
G298.862-0.438	&50.85$^{+0.12}_{-0.14}$   & 119  $^{+ 22}_{- 13}$  &11.15$^{+1.18}_{-1.37}$ &51.09$^{+0.21}_{-0.17}$	&Yes      \\
G305.359+0.194	&49.97$^{+0.19}_{-0.22}$   & 279  $^{+ 58}_{- 47}$  & 6.88$^{+0.24}_{-0.16}$ &50.32$^{+0.24}_{-0.35}$	&Likely   \\
G319.158-0.398	&50.17$^{+0.06}_{-0.05}$   &  34  $^{+  4}_{-  4}$  & 7.39$^{+0.22}_{-0.29}$ &50.43$^{+0.12}_{-0.17}$	&Yes      \\
G319.392-0.009	&50.10$^{+0.05}_{-0.07}$   &  66  $^{+  9}_{-  9}$  & 7.74$^{+0.20}_{-0.34}$ &50.32$^{+0.17}_{-0.14}$	&Yes      \\
G320.327-0.184	&47.67$^{+0.33}_{-0.48}$   & 116  $^{+ 43}_{- 24}$  & 7.84$^{+0.27}_{-0.30}$ &47.73$^{+0.30}_{-0.58}$	&No       \\
RCW97			&49.85$^{+0.07}_{-0.13}$   & 441  $^{+ 66}_{- 48}$  & 6.10$^{+0.26}_{-0.25}$ &50.16$^{+0.24}_{-0.18}$	&Likely      \\
G327.993-0.100	&48.70$^{+0.12}_{-0.10}$   & 160  $^{+ 20}_{- 23}$  & 6.14$^{+0.27}_{-0.26}$ &48.84$^{+0.17}_{-0.15}$	&No       \\
G330.868-0.365	&49.35$^{+0.11}_{-0.12}$   & 117  $^{+ 18}_{- 14}$  & 5.53$^{+0.33}_{-0.31}$ &49.58$^{+0.15}_{-0.20}$	&No       \\
G331.324-0.348	&48.99$^{+0.16}_{-0.16}$   & 256  $^{+ 38}_{- 31}$  & 5.64$^{+0.47}_{-0.41}$ &49.19$^{+0.23}_{-0.21}$	&No       \\
G331.354+1.072	&49.20$^{+0.11}_{-0.08}$   &  51  $^{+  8}_{-  6}$  & 4.87$^{+0.23}_{-0.34}$ &49.38$^{+0.16}_{-0.14}$	&No       \\
G331.529-0.084	&50.48$^{+0.24}_{-0.29}$   & 153  $^{+ 31}_{- 30}$  & 4.03$^{+0.54}_{-0.15}$ &50.97$^{+0.30}_{-0.44}$	&Yes      \\
G333.122-0.446	&49.56$^{+0.09}_{-0.07}$   & 214  $^{+ 32}_{- 20}$  & 6.13$^{+0.23}_{-0.22}$ &49.84$^{+0.18}_{-0.16}$	&Unlikely \\
G333.293-0.382	&49.56$^{+0.22}_{-0.25}$   & 266  $^{+ 58}_{- 46}$  & 6.11$^{+0.57}_{-0.57}$ &49.87$^{+0.27}_{-0.37}$	&Unlikely \\
G333.610-0.217	&49.99$^{+0.19}_{-0.30}$   & 451  $^{+ 97}_{- 82}$  & 6.18$^{+0.55}_{-0.60}$ &50.30$^{+0.35}_{-0.35}$	&Likely   \\
G338.398+0.164	&50.68$^{+0.04}_{-0.06}$   &  52  $^{+  6}_{-  6}$  & 6.30$^{+0.27}_{-0.38}$ &51.04$^{+0.16}_{-0.18}$	&Yes      \\
G338.400-0.201	&50.17$^{+0.05}_{-0.06}$   &  54  $^{+  8}_{-  7}$  & 8.47$^{+0.61}_{-0.34}$ &50.38$^{+0.16}_{-0.14}$	&Yes      \\
G345.555-0.043	&50.58$^{+0.06}_{-0.05}$   &  38  $^{+  4}_{-  4}$  & 7.49$^{+0.58}_{-0.35}$ &50.89$^{+0.14}_{-0.18}$	&Yes      \\
G345.645+0.009	&50.39$^{+0.05}_{-0.05}$   &  50  $^{+  6}_{-  5}$  & 7.10$^{+0.47}_{-0.38}$ &50.68$^{+0.15}_{-0.16}$	&Yes      \\
G347.611+0.204	&50.28$^{+0.10}_{-0.11}$   &  50  $^{+  8}_{-  6}$  & 1.80$^{+0.20}_{-0.06}$ &50.67$^{+0.16}_{-0.19}$	&Yes      \\
G351.467-0.462	&48.69$^{+0.11}_{-0.08}$   & 132  $^{+ 21}_{- 14}$  & 5.19$^{+0.28}_{-0.39}$ &48.88$^{+0.13}_{-0.16}$	&No       \\
Sgr C			&50.09$^{+0.05}_{-0.05}$   &  87  $^{+ 12}_{-  8}$  & 0.09$^{+0.18}_{-0.00}$ &50.51$^{+0.14}_{-0.14}$	&Yes 
\enddata
\end{deluxetable*}

\subsection{Our Updated List of G\ion{H}{2} Regions}\label{sec:list}

\begin{figure}[thb!]
\begin{center}
\includegraphics[width=2.6in]{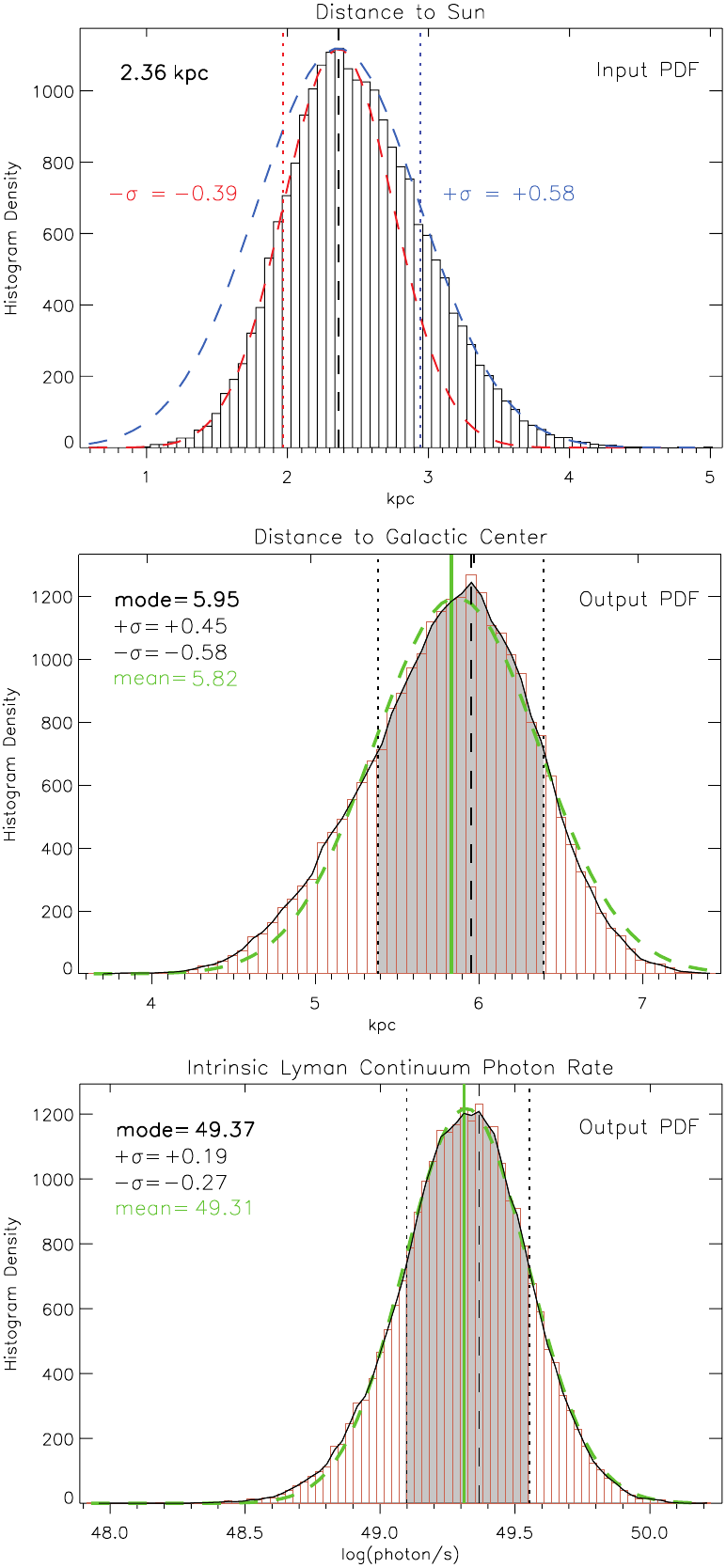}\\
\end{center}
\caption{\scriptsize Input and output probability distribution functions (PDFs) for select values associated with the Monte Carlo calculations for Sgr\,D. The top plot demonstrates graphically how the input PDFs were created for values with asymmetric uncertainties, in this case for the the source distance. The dashed red curve shows the Gaussian fit to the lower uncertainty and blue dashed curve for the upper uncertainty. The black dashed vertical line is the value for the distance to Sgr\,D, and the red and blue dotted vertical lines show the lower and upper uncertainty bounds, respectively. The bottom two plots show two example output (posterior) PDFs, for distance to Galactic center ($R_{GC}$) and intrinsic Lyman continuum photon rate ($N_{LyC}$). The black line is the fit to the histogram of the PDF, the black dashed vertical line shows the mode of the PDF. The shaded area shows the 68.2\% confidence interval, and the upper and lower bounds of this area are given by the dotted vertical lines (which represent the upper and lower uncertainties, respectively). The mode values and these upper and lower uncertainties for all sources and all output parameter PDFs are given in Table~\ref{tb:outputs}. For comparison, the green dashed lines are simple Gaussian fits to the black curves, and the vertical green lines show the mean values calculated from those fit.}
\label{fig:pdfs}
\end{figure}

We created probability distribution functions (PDFs) based upon the values and errors for all input variables given in Table~\ref{tb:inputs}, as well as for the constants $R_{0}$ and $a_o$, and the intermediary variable $x\sigma_{He}/x_1\sigma_{\nu}$. For input values with the same upper and lower uncertainties (i.e., reported with a standard deviation), the PDFs were created using a normal distribution (with width determined by the standard deviation). For values with asymmetric uncertainties, we created a PDF from two half-gaussians whose width is determined by the two uncertainties and normalized to the same peak. We created a Monte Carlo code that chose values randomly from each of these PDFs and used them as the inputs for the equations in Section~\ref{sec:lyman} and repeated this 25000 times. From the posterior PDFs generated by this process, the most likely (mode) values given in Table~\ref{tb:outputs} are listed for the observed Lyman continuum photon rate ($N^{\prime}_{LyC}$), electron density ($N_e$), galactocentric distance ($R_{GC}$), and intrinsic Lyman continuum photon rate ($N_{LyC}$) for all sources. These posterior PDFs for the above parameters were almost always skewed somewhat, so in order to calculate the uncertainties in the mode values, we determined the lower and upper bounds of the PDF such that the value of the PDF was equal at both bounds and the area under the PDF between the two bounds was equal to 68.2\% of the area under the total PDF. This is analogous to the standard deviation of a normal distribution which also represents a 68.2\% confidence interval. To help visualize this, we provide in Figure~\ref{fig:pdfs} graphs of one of the input distributions with asymmetric errors (distance to the Sun), as well as two example output distributions ($R_{GC}$ and $N_{LyC}$) with the mode, mean, and calculated 68.2\% confidence intervals shown. These uncertainties are reported for all values given in Table~\ref{tb:outputs}.

We preserve the input values from \citet{2004MNRAS.355..899C} for 6\,cm flux density ($S_{6cm}$; Table~\ref{tb:inputs}, column 8) and 6\,cm source size ($\theta_{6cm}$; Table~\ref{tb:inputs}, column 9) since the surveys that performed these observations are from the single-dish radio antennas with the ability to resolve sources equal to or greater than $\sim$2$\arcmin$, which is adequate for the typical sizes of G\ion{H}{2} regions. More recent radio continuum observations have been taken of several of these G\ion{H}{2} regions, many with sub-arcsecond resolutions, however these observations are taken with interferometric arrays and thus are not as good for observing extended large-scale emission which is filtered out to varying degrees. 

Generally, even distances that are determined with extreme accuracy via say, maser parallaxes, are within reasonable agreement with either the near or far kinematic distance. As Sgr\,D and W42 show us, the largest changes in the calculation of $N_{LyC}$ is when a source distance is changed from a far distance to near. Therefore the most impactful measurements since \citet{2004MNRAS.355..899C}, and those that will change the G\ion{H}{2} region census the most, will be those that help to resolve near/far kinematic distance ambiguities by changing the accepted distance from far to near. That being said, however, there are a good number of sources in our updated list that fall relatively close to the $N_{LyC} = 10^{50}$~photons/s criterion, so a proper estimate of the errors for each source was warranted in order to clearly indicate how likely a source is or is not a G\ion{H}{2}. 

One additional caveat is that we assumed in Section~\ref{sec:lyman} a $R_0$ of 8.34$\pm$0.16\,kpc, however the IAU recommended value is 8.5\,kpc, and there exist recent measurements that place the galactic center as close as 7.9\,kpc \citep{2020PASJ...72...50V}. Using an inaccurate $R_0$ value does not affect the calculations of $N^{\prime}_{LyC}$ or $N_e$, and only affects the calculations of $R_{GC}$ and $N_{LyC}$. The difference in these values when one assumes $R_0=8.5$\,kpc versus 7.9\,kpc, is up to $\pm$0.6\,kpc in the calculation of $R_{GC}$ and $\pm$0.03\,dex in the calculation of log$N_{LyC}$.

\citet{2004MNRAS.355..899C} define a G\ion{H}{2} as a source with $N_{LyC} >10^{50}$~photons/s and we find that with our new calculations that many sources no longer satisfy that criterion (see the right-most column of Table~\ref{tb:outputs}). Of the 56 sources in the original census, we find that 12 can no longer be considered G\ion{H}{2} regions. An additional two sources have $N_{LyC}$ values below the $10^{50}$~photons/s criterion, but their upper limit errors do cross over this cut-off value. These sources are considered ``Unlikely'' to be G\ion{H}{2} regions in Table~\ref{tb:outputs}. There are a further eleven regions where their $N_{LyC}$ values are above the cut-off criterion, but the lower limits errors do go below the cut-off value. These sources are considered ``Likely'' to be G\ion{H}{2} regions in Table~\ref{tb:outputs}. 

Related to this cut-off value, however, is the question of whether or not the G\ion{H}{2} criterion of log$N_{LyC} > 50.0$~photons/s should be a strict or loose criterion. \citet{2004MNRAS.355..899C} claim that this number was chosen because it was close to the equivalent of 10 O7\RNum{5} stars. Whether or not this is true depends on the stellar models used \citep[e.g., this is about 14 O7V stars according to][]{1973AJ.....78..929P}. Giant \ion{H}{2} regions were first defined, in an admittedly arbitrary fashion, by \citet{1970IAUS...38..107M} as having $S_{5}d^2 > 400$, with $S_{5}$ being the 5\,GHz (i.e., 6\,cm) radio flux density in Jy, and $d$ the distance to the source in kpc. Plugging 400 in for the last two terms in Equation~\ref{equ:Npc} and assuming $T_e=10000$\,K yields the criterion used by \citet{1978AA....66...65S} of G\ion{H}{2} regions requiring log$N^{\prime}_{LyC} > 49.6$ (i.e., defined by the observed rather than intrinsic Lyman continuum photon rate). Looking to the log$N^{\prime}_{LyC}$ values we derived for the sources in Table~\ref{tb:outputs}, it can be seen that all sources with log$N^{\prime}_{LyC} > 49.6$ also have log$N_{LyC} > 50.0$, and thus it would seem that setting the criteria based upon intrinsic or observed Lyman continuum photon rate does not greatly change which sources are considered to be, or not to be, G\ion{H}{2} regions. 

There are also an additional three sources with significant unresolved ambiguity. As pointed out in Table~\ref{tb:kinematic} and in Appendix~\ref{sec:app}, G320.327-0.184 and G338.398+0.164 have kinematic distances with conflicting absorption line measurements pointing to both near and far distances. G320.327-0.184 was chosen to be at the near distance because the \ion{H}{1} observations are more recent. G338.398+0.164 has multiple ATLASGAL sub-mm clumps within the radio emission region with different indicated near/far distances from \ion{H}{1} absorption measurements, but the central source has \ion{H}{1} absorption indicating the far distance. G338.400-0.201 is uncertain because there is a wide range of measured $v_{lsr}$ values, some of which indicate the region is in the far outer Galaxy, however the infrared component as seen by MSX is compact and relatively faint, which is highly unusual for a G\ion{H}{2} region. Nonetheless, we keep the far distance in keeping with previous studies.

To summarize, 42 of the original census of 56 G\ion{H}{2} regions appear to be, or are likely to be, G\ion{H}{2} regions. This means that 25\% of the original census are below the cut-off log$N_{LyC}$ value to be considered being bona fide G\ion{H}{2} regions. Furthermore, another 20\% of the original census have errors that dip below the cut-off value, so their status as a bona fide G\ion{H}{2} is less certain. We stress that while the original census of G\ion{H}{2} regions compiled by \citet{2004MNRAS.355..899C} was the most extensive vetted list available, it is not considered a complete list of all radio G\ion{H}{2} regions in the Milky Way. While it would be interesting to do a more thorough compilation of all G\ion{H}{2} sources (including revisiting the sources originally rejected by \citealt{2004MNRAS.355..899C}), this census (even our new pared-down version) does contain the brightest and most well-known sources, and contains a sufficient number of sources for the purposes of our SOFIA survey.

\section{Discussion}\label{sec:reject}

As discussed in \citet{2004MNRAS.355..899C} and the introduction to this paper, apart from having Lyman continuum photon rates in excess of $N_{LyC} = 10^{50}$\,photons/s, G\ion{H}{2} regions are also considered to be the formation sites of our Galaxy's most massive OB clusters. Certainly, the highest values of $N_{LyC}$ observed in some G\ion{H}{2} regions (e.g., NGC\,3603) cannot physically be due to a single O-type star, and must be due to a sizeable cluster or multiple generations of clusters of massive, ionizing stars. However, as one goes to lower values of $N_{LyC}$ it can become unclear if the region is predominantly powered by a single, very massive star or a cluster of less-massive ionizing stars. Generally speaking, as we go to earlier and earlier O-star spectral types, they rapidly become increasingly rare, and therefore the derived $N_{LyC}$ value for modestly powerful \ion{H}{2} regions becomes less likely to be due to a single, powerful O-star and more likely to be due to a cluster of stars with the combined ionizing rate equal to that same $N_{LyC}$ value. However, this is not always going to be the case, as we see in the examples of W42 and Sgr\,D in this work where they appear to be predominantly ionized by a single source. Therefore, because all \ion{H}{2} regions lie on a continuum of $N_{LyC}$ values, any cut-off value is arbitrary and alone may not be sufficient to ascertain whether a region's energy is dominated by a very powerful ionizing source, or a large cluster of less-powerful (but still massive) ionizing sources (i.e., a large OB cluster or protocluster). 

Another example of this comes from looking at the sources rejected as G\ion{H}{2} regions in Table~\ref{tb:outputs}, where we see that one of the sources with the smallest measured log$N_{LyC}$ is G351.467-0.462, with a value of 48.88 photons/s. This is the equivalent of a single main sequence O7 star of $\sim30\,M_{\sun}$. This, of course, does not by itself mean that there is definitively only one O7 star ionizing G351.467-0.462. In principal, however, G351.467-0.462 could indeed be a \ion{H}{2} region powered by a single O7 ionizing star, rather than a star-forming region producing an OB stellar cluster (with, e.g., $\sim$30 B0 stars which is also equivalent to log$N_{LyC} = 48.88$\,photons/s). A well-known analog would be $\theta^1$ Orionis C1, which is an O6\RNum{5} star \citep{2014AstBu..69...46B} that is singularly responsible for generating the vast majority of the ionizing photons in the Orion Nebula \citep[log$N_{LyC}$$\, = 49.47$ photon/s;][]{2001ApJ...555..613I}. Though the Orion Nebula is impressive due to its proximity, there is a stark contrast between Orion and the most powerful object in the Milky Way, NGC\,3603. By way of comparison, NGC\,3603 has the equivalent of $\sim$100 times the $N_{LyC}$ of Orion, and 10 times the number of revealed O stars \citep[i.e., 50 stars $\geq$15$\,M_{\sun}$ compared to 5;][]{1998ApJ...498..278E,1997AJ....113.1733H}. Also by way of comparison, the $N_{LyC} = 10^{50}$\,photons/s cut-off is the equivalent to four times that of Orion or the equivalent of single O4 ZAMS star \citep{1973AJ.....78..929P} with a mass of 65\,$M_{\sun}$ \citep{2000AJ....119.1860B}.

We can look at the multi-wavelength observations of Sgr\,D and W42 and see if there is any supporting evidence (beyond just $N_{LyC}$) that would demonstrate that these two regions, in particular, are not likely to be G\ion{H}{2} regions. Indeed, qualitatively Sgr\,D and W42 have much simpler morphologies compared to our previously observed G\ion{H}{2} regions of this project: G49.5-0.4 and G49.4-0.3 in W51A \citepalias{2019ApJ...873...51L}, M17 \citep{2020ApJ...888...98L}, and W49A \citep{2021ApJ...923..198D}. The radio and MIR emission from both Sgr\,D and W42 are dominated by a single, bright and relatively compact region. Besides the extended source associated with the main infrared peak, Sgr\,D has only two other extended (but diffuse) infrared sources (A and source 2), while W42 has no other separate, extended infrared sources in its vicinity. This can be contrasted with the previously studied G\ion{H}{2} regions which are broken up into multiple, bright, extended, and separate star-forming sub-regions like G49.5-0.4 and G49.4-0.3 in W51A (10 and 5 sub-regions, respectively), W49A (15 sub-regions), or M17 (4 sub-regions). Besides the extended mid-infrared emission sources associated with Sgr\,D and W42, there are 3 compact mid-infrared sources in Sgr\,D (source B, C, and D) and 2 in W42 (W42-MME and source 1). This is far fewer than the number of compact sources seen in our previously studied G\ion{H}{2} regions G49.5-0.4 (37), G49.4-0.3 (10), W49A (24), and M17 (16) which lie at a wide range of distances (i.e., $\sim$2$-11$\,kpc). The dearth of compact sources within Sgr\,D and W42 would seem to imply less-vigorous star formation activity (regardless of their distances) than what we see in the G\ion{H}{2} regions studied thus far. Furthermore, the best fit mass estimates of the most massive YSOs in Sgr\,D and W42 (Table~\ref{tb:sed}) is only 16 and 32\,$M_{\sun}$, respectively, which is more modest compared to the largest best fit masses of the MYSOs seen in M17 (64\,$M_{\sun}$), W49A (128\,$M_{\sun}$), G49.5-0.4 (96\,$M_{\sun}$), and G49.4-0.3 (64\,$M_{\sun}$). 

If Sgr\,D and W42 have such low $N_{LyC}$, we might suspect that they are likely to be like Orion and have a single O star responsible for the majority (if not all) of their emission. In fact, in the case of W42 there is a confirmed O star seen in the near-infrared that has the equivalent $N_{LyC}$ of the entire \ion{H}{2} region and is therefore overwhelmingly responsible for ionizing the entire region \citep{2000AJ....119.1860B}. A \ion{H}{2} region predominantly ionized by a single star would also mean that the majority of the associated mid-infrared dust emission should be concentrated around a single peak, as is the case for Sgr\,D and W42, rather than coming from multiple, separate, star-forming clumps. Quantitatively, the percentage of the flux in the brightest peak compared to the total flux in the whole \ion{H}{2} region is $\sim$85\% for Sgr\,D and $\sim$50\% for W42 at 37\,$\mu$m. This can be compared to the G\ion{H}{2} regions we have already studied that have more dispersed flux spread throughout their entire volume and/or broken up into multiple star-forming sub-regions. The percentage of the flux in the brightest peak to the total flux in the whole G\ion{H}{2} region is $\sim$20\% for G49.5-0.04, $\sim$15\% for G49.4-0.03, $\sim$25\% for W49A, and $\sim$5\% for M17 at 37\,$\mu$m.  

In summary, the case studies of Sgr\,D and W42 reveal that they have multi-wavelength properties that are in keeping with the idea that they are more modest \ion{H}{2} regions accompanied by less-vigorous star formation activity, rather than being like our previously studied G\ion{H}{2} regions (i.e., G49.5-0.4 and G49.4-0.3 in W51A, M17, and W49A). In particular, since the majority of the infrared (and/or radio) flux in the entire \ion{H}{2} region comes from a single, compact region in both these objects, it indicates that they are predominantly powered by a single ionizing source (similar to Orion) rather than a massive OB cluster or protocluster. For some sources with large errors in their derived $N_{LyC}$ values or large uncertainties in their distances, one could potentially use similar analyses to be able to differentiate between sources likely to be (or not to be) G\ion{H}{2} regions. However, we caution that this proposed distinction between G\ion{H}{2} and \ion{H}{2} regions is only based upon a small number of sources (i.e., four G\ion{H}{2} regions and two \ion{H}{2} regions), and G\ion{H}{2} and \ion{H}{2} regions likely have some variety of properties that cannot be accounted for with such small numbers. As we continue our survey of G\ion{H}{2} regions, we will address the observational properties as a function of $N_{LyC}$ in a later paper when there are more sources to be of statistical significance to make more nuanced conclusions about the G\ion{H}{2} and \ion{H}{2} populations as a whole. 

\section{Summary}\label{sec:sum}

In the first part of this paper we present the new SOFIA infrared imaging data that we obtained for Sgr\,D at 20 and 37\,$\mu$m and W42 at 25 and 37\,$\mu$m. We discuss how the updated, nearer distances measured towards both Sgr\,D and W42 disqualify them as being bona fide G\ion{H}{2} regions. Nonetheless, we derive and discuss the detailed physical properties of the individual compact sources and sub-regions as well as the large-scale properties of the two star-forming regions based upon the SOFIA data and other multi-wavelength data. 

While the radio region of Sgr\,D is a fairly circular 6.6$\arcmin$ diameter \ion{H}{2} region, bright mid and far-infrared emission only comes from a small number of discrete locations. For Sgr\,D, we suggest that the three brightest mid-infrared sources, sources 2, 3, and D, are all coincident with (and likely formed out of) a dark filament induced by the collision of the Sgr\,D \ion{H}{2} region with supernova remnant G1.05-0.15. Within Sgr\,D we find only three MYSO candidates. The brightest mid-infrared source, source 3, appears to be at least partially embedded in the dark filament and has a infrared morphology similar to an edge-on flared disk.  

Our SOFIA images of W42 at 25 and 37\,$\mu$m detect a single, extended emission region with an extent similar to that seen by MSX at 22\,$\mu$m, i.e., about 2$\arcmin$ in diameter. The central 30$\arcsec$ region of W42 has a bright mid-infrared peak coincident with a known radio compact \ion{H}{2} region (G025.3824-00.1812), a second peak coincident with methanol maser emission (a tracer of high-mass YSOs), and a third peak to the north associated with fainter radio continuum emission. However, there is an O5-O6.5 star (W42\#1) near the center of W42 \citep{2000AJ....119.1860B} whose Lyman continuum photon rate alone is equal to that of the entirety of W42. Due to its more evolved nature, it apparently does not have much circumstellar dust since it has no detectable mid-infrared emission of its own as seen by SOFIA. It is unclear how much of the radio and infrared emission attributed to the sources associated with the infrared and radio peaks may actually be due to external ionization and heating by W42\#1, but it does appear that W42\#1 may be solely responsible for the ionization of the vast majority of W42. Our SED modeling shows three massive YSOs may be present here, however this assumes no contamination from W42\#1, which may not be the case. 

In the second part of this paper we compiled data that updated the distances to the census of 56 G\ion{H}{2} regions identified by \citet{2004MNRAS.355..899C}. We recalculated their Lyman continuum photon rates, $N_{LyC}$, and determined that 25\% of these sources (14) are at sufficiently closer distances that their derived values of $N_{LyC}$ are $<10^{50}$\,photons/s, meaning they no longer meet the criterion to be considered G\ion{H}{2} regions in the strictest sense. Of the remaining 42 G\ion{H}{2} region candidates identified here, an additional 20\% (11) have $N_{LyC}>10^{50}$\,photons/s but have measurement errors that could place them below the cut-off value for being bona fide G\ion{H}{2} regions. 

We additionally looked at other observational and physical characteristics (besides Lyman continuum photon rate) of Sgr\,D and W42 and compared these properties to those of the G\ion{H}{2} regions that we have already studied as part of the SOFIA G\ion{H}{2} region survey. We determine that Sgr\,D and W42 appear to have much simpler morphologies in the infrared, seem to have a dearth of compact infrared sources, and have observational characteristics that indicate that they are dominantly ionized by single massive stars and not large OB clusters. Additionally, the most massive MYSOs in Sgr\,D and W42 are only 16 and $32\,M_{\sun}$, respectively, while our previous observations of the brightest G\ion{H}{2} regions have most massive MYSOs in the range of $64-128\,M_{\sun}$. Given that the Lyman continuum photon rate for a single \ion{H}{2} region exists within a continuum of $N_{LyC}$ values provided by the population of all \ion{H}{2} regions, any cut-off value is somewhat arbitrary. We suggest, based upon what we have learned from Sgr\,D and W42, that if a source has a derived value near the $N_{LyC}$ cut-off and/or has large uncertainty in this value or its distance, other observational indicators such as those described above could potentially be used to help determine if it should be disqualified as a bona fide G\ion{H}{2} region.

\begin{acknowledgments}
This research is based on observations made with the NASA/DLR Stratospheric Observatory for Infrared Astronomy (SOFIA). SOFIA is jointly operated by the Universities Space Research Association, Inc. (USRA), under NASA contract NAS2-97001, and the Deutsches SOFIA Institut (DSI) under DLR contract 50 OK 0901 to the University of Stuttgart. This work is also based in part on archival data obtained with the Spitzer Space Telescope, which is operated by the Jet Propulsion Laboratory, California Institute of Technology under a contract with NASA. This work is also based in part on archival data obtained with Herschel, an European Space Agency (ESA) space observatory with science instruments provided by European-led Principal Investigator consortia and with important participation from NASA.
\end{acknowledgments}

\facility{SOFIA(FORCAST)}

\appendix

\section{Discussion of Distance Determinations for Each Region}\label{sec:app}

\textbf{G0.361-0.780/G0.394-0.540/G0.489-0.668/G0.572-0.628:} These four sources all lie within a 10$\arcmin$ area centered at ($\ell$,~$b$)=(0.45,~-0.67), in a dusty complex that lies at a similar Galactic longitude as Sgr\,B but more than a half a degree below the plane of the central Galaxy described by Sgr~B, Sgr~A, and Sgr~C. For these sources there exist kinematic distance determinations only, and in and around the Galactic center ($357.5 < \ell < 2.5$) such measurements are not reliable because the sources have more complex orbits than those for sources that lie in the Galactic disk at larger galactocentric radii \citep{2018ApJ...856...52W}. Thus, in many studies such Galactic Center sources are simply assumed to be the same distance as Sgr\,A\text{*}. However, as we see in the case of Sgr\,D this is not always an accurate assumption. Observations of H110$\alpha$ by \citet{1980AAS...40..379D} toward each of these sources yields v$_{lsr}$ values between 17 and 24\,km/s, for which we calculate near kinematic distances between 7.66 and 8.0\,kpc; far distances are nearly the same and range from 8.30 to 8.48\,kpc. The only one of these sources with observed absorption features is G0.572-0.628, however \citet{2013ApJ...774..117J} state that the observed \ion{H}{1} absorption seen towards this source does not give conclusive evidence for resolving the distance ambiguity. \citet{2014AA...569A.125H} adopt the near distances to all of these sources and reference \citet{2003AA...397..133R} as the source of the disambiguation. \citet{2003AA...397..133R} state that the near distances are preferred as these sources exhibit an H$\alpha$ counterpart. We will assume these near distances here as well. 

\textbf{Sgr\,B1:} The source closest to the radio region galactic coordinates ($\ell$ = 0.518, $b$ = -0.065) given by \citet{2004MNRAS.355..899C} is Sgr\,B1 (which is coincident to within an arcminute). However, these coordinates were identified as W24 by \citet{2004MNRAS.355..899C}, which is usually synonymous with Sgr\,B2, though that source is 9$\arcmin$ away from Sgr\,B1. Though these separations seem quite large, both Sgr\,B1 and Sgr\,B2 are believed to be part of the same molecular cloud \citep{2021ApJ...910...59S} and have similar v$_{lsr}$ velocities. The distance to Sgr\,B2 was determined via maser parallax to be $7.8^{+0.8}_{-0.7}$\,kpc, and it is assumed here that this distance applies to Sgr\,B1 as well.

\textbf{G2.303+0.243:} Because OH absorption lines are found with velocities ($\sim$16\,km/s) greater than those of the source (5\,km/s), \citet{2003AA...397..133R} adopt the far distance to this source. Using the the most precise velocity measurement, v$_{lsr}$=4.9$\pm$0.7\,km/s, which is the H87$\alpha$+H88$\alpha$ transition from \cite{1989ApJS...71..469L}, we derive a kinematic far distance of 13.48\,kpc.

\textbf{G3.270-0.101:} The H110$\alpha$ measurement of \citet{1980AAS...40..379D} toward this source has a v$_{lsr}$ value of 4.3$\pm$5.0\,kpc, however we will use the comparable but more precise measurement in the H87$\alpha$+H88$\alpha$ transitions from \cite{1989ApJS...71..469L} of v$_{lsr}$=4.9$\pm$0.8\,km/s. This velocity yields a far kinematic distance of 14.33\,kpc which we adopt here given that \citet{2013ApJ...774..117J} state that there are \ion{H}{1} absorption features seen at  velocities corresponding to both the Near and Far 3 kpc Arms, suggesting a distance at least as far as the Far 3 kpc Arm.

\textbf{G4.412+0.118:} This is a source has a 6\,cm radio diameter of almost 5$\arcmin$, but there are multiple molecular clumps within even 2$\arcmin$ of the galactic coordinates of this source. The closest source in the ATLASGAL catalogue \citep{2016PASA...33...30R} is 0.6$\arcmin$ away and has a N$_2$H$^+$ velocity of 4.23$\pm$0.25\,km/s, however there are two other sources within 2$\arcmin$ of this location with values of 3.0 and 8.5\,km/s. The H110$\alpha$ measurement of \citet{1980AAS...40..379D} toward this source has a v$_{lsr}$ value of 5.7$\pm$5.0\,km/s, and the more precise observations of the H87$\alpha$+H88$\alpha$ transitions from \cite{1989ApJS...71..469L} give v$_{lsr}$=4.1$\pm$0.9\,km/s. \citet{2013ApJ...774..117J} claim this source is likely at the far kinematic distance due to \ion{H}{1} absorption present at multiple velocities, and quote distance (with large errors) of 15.6$\pm$8.9\,kpc. Using our adopted value of 4.23\,km/s (chosen for it's precision) we calculate a far distance of $14.97^{+0.77}_{-0.58}$\,kpc. The far distance is also suggested by \citet{2003AA...397..133R} due to absorption features seen at velocities greater than that of the source.

\textbf{M8:} The v$_{lsr}$ of this source has been observed by multiple groups in multiple transitions in the past 50 years. Choosing from one of the papers we have cited previously in this work, \cite{1989ApJS...71..469L} measured a v$_{lsr}$=4.1$\pm$0.9\,km/s in the H87$\alpha$+H88$\alpha$ transition, and the source was believed to be at the near kinematic distance due to its low galactic latitude (i.e., -1.178$\arcdeg$). \citet{2011MNRAS.411..705M} claim that their spectrophotometry of sources within M8 agree with the near kinematic distance of 2.8\,kpc claimed by \citet{2003AA...397..133R}. However, SED fitting to near-infrared sources in M8 performed by \citet{2006MNRAS.366..739A} have estimated an even closer distance of 1.25\,kpc. Indeed the  most recent measurements made using GAIA parallaxes of cluster members within M8 support this closer distance, for instance \citet{2018ApJ...864..136B} measuring 1.17$\pm$0.10\,kpc, \citet{2019AA...623A..25D} measuring 1.325$\pm$0.113\,kpc, as well as our adopted value of 1.34$\pm$0.07\,kpc measured by \citet{2020AA...633A.155R}.

\textbf{G8.137+0.228}: There have been many observations of the v$_{lsr}$ of this source in multiple hydrogen transitions as well as CS, leading to values between 19.3 and 24.4\,km/s \citep[see][]{2014AA...569A.125H}. We adopt the velocity measurement with the smallest error of v$_{lsr}$=20.31$\pm$0.06 from \citet{2006ApJ...653.1226Q}. \citet{2004MNRAS.355..899C} adopted the far distance to this source, however most recent studies prefer the closer distance \citep[e.g.,][]{2019ApJ...878...26D, 2018MNRAS.473.1059U}. \citet{2013ApJ...774..117J} claim that a near-side kinematic value is preferred because they do not see any \ion{H}{1} absorption features. 

\textbf{W31:} W31 consists of multiple extended H II regions, of which G10.315-0.150 (W31-North), G10.159-0.349 (W31-South), and G10.62-0.38 are the brightest in cm radio continuum emission. A distance determination to W31 is very complicated. For W31-South, \citet{2004MNRAS.355..899C} give a distance of 4.5\,kpc that is from \citet{2004AA...419..191C}. That work claims that because several measurements have shown absorption lines observed up to 43\,km/s \citep[e.g.,][]{1974AA....31...83W}, a distance less than $\sim$2.5\,kpc derived by some kinematic studies would not make sense, and that at 43\,km/s the source would exist at 4.5\,kpc. However, from \citet{2018MNRAS.473.1059U} it can be seen that there are five ATLASGAL clumps within 2$\arcmin$ of these galactic coordinates (and the source has a measured 6\,cm radio diameter of 2.9$\arcmin$). The closest clump ($\sim$0.6$\arcmin$ away) has a v$_{lsr}$ of 8.9 km/s, while all the others have v$_{lsr}$ of 13.0--14.4\,km/s. Therefore, there are likely to be multiple clumps present at different distances along the line of sight to this source that are confusing the issue. 
Contrary to \citet{2004AA...419..191C}, \citet{2018MNRAS.473.1059U} claim that the \ion{H}{1} self-absorption observations argue for a near kinematic value for the closest ATLASGAL clump and give a kinematic distance of 1.3\,kpc.  

Luckily, the distance to this source has been measured by what is generally considered a more accurate method. That is, there have been two spectrophotometric distance measurements towards the NIR-bright stars within W31-South, the first by \citet{2001AJ....121.3149B} who derive a spectrophotometric distance of 3.4$\pm$0.3\,kpc, and the second by \citet{2011MNRAS.411..705M} who derive a distance of 3.55$\pm$0.94\,kpc. We will adopt here the more accurate spectrophotometric distance of \citet{2001AJ....121.3149B}. As we discuss in Section~\ref{sec:dist}, the fact that this value does not match either the kinematic distance, nor the \ion{H}{1} absorption velocities, is not uncommon.  

Similarly for W31-North, \citet{2015AA...582A...1D} discuss in-depth the multitude of conflicting distance measurements that lead to a distance range of 2 to 19\,kpc, but argue based upon their spectrophotometric analysis that the region lies at a distance of 1.75$\pm$0.25\,kpc, which is compatible with the near kinematic distance. For W31-North, we again adopt the spectrophotometric distance due to the fact that the methodology is more accurate than the kinematically-derived distances. 

Though it is not in our source list, the third major region within W31, G10.62-0.38, has had accurate maser parallax measurements performed by \citet{2014ApJ...781..108S}, placing it at $4.95^{+0.51}_{-0.43}$\,kpc, which they claim is the distance to the entirety of W31. Given that the source complexity within this region and the fact that G10.62-0.38 is almost a half-degree from either W31-South or W31-North, we do not assign this maser distance to either source. 

\textbf{M17:} There are two maser parallax measurements towards this region, consistent with each other to within the errors, one by \citet{2011ApJ...733...25X} who measured $1.98^{+0.14}_{-0.12}$\,kpc and one by \citet{2016MNRAS.460.1839C} who measured $2.04^{+0.16}_{-0.17}$\,kpc. We studied M17 in-depth in \citetalias{2020ApJ...888...98L} and adopted the value from \citet{2011ApJ...733...25X} which had slightly smaller errors. These values are also consistent with the near kinematic distances measured by multiple studies, for instance the H87$\alpha$+H88$\alpha$ transition measurements from \cite{1989ApJS...71..469L}, which have a quoted v$_{lsr}$=16.8$\pm$0.3\,km/s, yields a kinematic near distance of $1.97^{+0.15}_{-0.38}$\,kpc.

\textbf{G20.733-0.087:} This source only has kinematic measurements, and the v$_{lsr}$ values all seem to hover in the 55.6--59.0\,km/s range. There are four ATLASGAL clumps within 2$\arcmin$ of these galactic coordinates \citep{2018MNRAS.473.1059U}, three of which have v$_{lsr}$ values in this range as well. However, there is one clump more than 1.8$\arcmin$ from the galactic coordinates with a v$_{lsr}$= 103.3\,km/s. Consistent with this, \citet{2003AA...397..133R} detects a H$_{2}$CO transition here at 104\,km/s as well as 56\,km/s.  \citet{2018MNRAS.473.1059U} claim the far distance is more likely due to \ion{H}{1} self-absorption, and this far distance appears to be the general consensus \citep[e.g.,][]{2006ApJ...653.1226Q,2003AA...397..133R}. Adopting the v$_{lsr}$ of 55.96$\pm$0.04 of \citet{2006ApJ...653.1226Q} yields a kinematic far distance of $11.69^{+0.34}_{-0.44}$\,kpc.

\textbf{G29.944-0.042:} The MSX images of this region from \citet{2004MNRAS.355..899C} show a group of about six extended infrared sources all within a 4$\arcmin$ radius. \citet{2014ApJ...781...89Z} present the velocity integrated $^{13}$CO maps of this region from the Galactic Ring Survey \citep{2006ApJS..163..145J}, revealing a single structure that encompasses all of the mid-infrared MSX sources centered very close to the galactic coordinates of this region. \citet{2014ApJ...781...89Z} also measure the parallax to two separate methanol maser sources within this $^{13}$CO clump. The maser source G029.95-00.01, lies just under 2$\arcmin$ from the galactic coordinates, and has a parallax that yields a distance of $5.26^{+0.62}_{-0.50}$\,kpc. The second maser source, G029.86-00.04, lies just over 5$\arcmin$ from the galactic coordinates, and has a parallax that yields a distance of $6.21^{+0.88}_{-0.69}$\,kpc. \citet{2014ApJ...781...89Z} further state that because the distances and proper motions of G029.86-00.04 and G029.95-00.01 are consistent to within their uncertainties, they are likely located at the same distance, and they compute a variance-weighted average distance of $5.71^{+0.50}_{-0.42}$\,kpc which we adopt here. Consistent with this logic is the fact that both maser sources lie in the same $^{13}$CO clump and that all of the ATLASGAL clumps within 8$\arcmin$ of these galactic coordinates have similar v$_{lsr}$ values in the range of 95--101\,km/s \citep{2018MNRAS.473.1059U}, which yield near kinematic distances of 5.7--6.2\,kpc (which is consistent with the maser-derived distance).  

\textbf{W43:} The distance to W43 was estimated by \citet{2014ApJ...781...89Z} based upon water and methanol maser parallaxes to two maser sources that lie approximately $-1.5\arcdeg$ in galactic longitude from W43 (i.e., the masers discussed above for G29.944-0.042), and two that lie approximately +1.5$\arcdeg$ from W43 in longitude. Given that the molecular clumps found within this entire 3$\arcdeg$ area have very similar v$_{lsr}$ values, it seems reasonable to assume that these masers which are not coincident with W43 still provide a good distance estimate to W43. \citet{2014ApJ...781...89Z} estimate $5.49^{+0.39}_{-0.34}$\,kpc based upon the average distance to the four methanol and water masers sources. We adopt this value in this work because it the same value to within the errors of the spectrophotometrically-derived distance from \citet{2011MNRAS.411..705M} of 4.90$\pm$1.91\,kpc. These values are also consistent to within the errors with the kinematically-derived near distances quoted for this region, for example the H91$\alpha$ transition velocity of 92.02$\pm$0.04\,km/s from \citet{2006ApJ...653.1226Q} which yields a near kinematic distance of $5.57^{+0.38}_{-0.73}$\,kpc.

\textbf{G32.8+0.19:} The distance to this source was derived from maser parallax observations of \citet{2019AJ....157..200Z}, which usually have accuracies of $\pm$1.0 kpc or less. However, this source distance is quoted with rather high errors. \citet{2019AJ....157..200Z} quote two distance estimates using two different methods of converting the parallax to distance: $9.7^{+4.1}_{-2.2}$\,kpc and $10.0^{+5.1}_{-2.7}$\,kpc. This makes its status as a G\ion{H}{2} unclear, since at 9.7\,kpc the source has log$N_{LyC}=49.90$\,photons/s, and at 9.7+4.1\,kpc it would have 50.15\,photons/s. These maser distances are consistent (within their errors) with the far kinematic distance of $12.85^{+0.44}_{-0.34}$\,kpc using a v$_{lsr}$ value of 15.46$\pm$0.15\,km/s from \citet{2006ApJ...653.1226Q} based on their measurements of the H91$\alpha$ transition. Given the high errors associated with the maser measurements, we will adopt this far kinematic distance in this work. \citet{2014AA...569A.125H} list multiple v$_{lsr}$ measurements toward this source from multiple groups, all of which have values of v$_{lsr}$=15.0--17.0\,km/s, but the measurement of \citet{2006ApJ...653.1226Q} has the smallest error.

\textbf{W49A:} This source was covered in detail in our \citetalias{2021ApJ...923..198D}. This source has reliable maser parallax measurements from \citet{2013ApJ...775...79Z} of $11.11^{+0.79}_{-0.69}$\,kpc. 

\textbf{G48.596+0.042:} There are two conflicting maser parallax measurements towards this region. \citet{2011PASJ...63..719N} measure a trigonometric parallax to the water masers in the source G48.61+0.02, which is $\sim$75$\arcsec$ from the galactic coordinates of this region and lies within the 6\,cm radio emitting diameter (i.e., 4.2$\arcmin$), deriving a distance of 5.03$\pm$0.19\,kpc. On the contrary, \citet{2013ApJ...775...79Z} claim a distance to this same source as $10.75^{+0.61}_{-0.55}$\,kpc also based upon water maser observations. \citet{2013ApJ...775...79Z} give multiple reasons to suspect the results of \citet{2011PASJ...63..719N}, but such large discrepancies are not common.  

The kinematic distance measurements towards this region all indicate a far distance of around 10\,kpc, for instance the H110$\alpha$ velocity from \citet{2002ApJS..138...63A} which is measured to be v$_{lsr}$=18.0$\pm$0.4\,km/s, which yields a far distance of $9.83^{+0.41}_{-0.44}$\,kpc. Since this value is consistent to within the errors of the maser measurement of \citet{2013ApJ...775...79Z}, we are inclined to believe it more and adopt it in this work. The near kinematic distance is only $1.21^{+0.35}_{-0.42}$\,kpc, which means the distance of \citet{2011PASJ...63..719N} is incompatible with both the near and far kinematic distances. Therefore, the only way that the distance value of \citet{2011PASJ...63..719N} could be right was if G48.61+0.02 had a very high peculiar velocity.

\textbf{G48.9-0.3:} This source is a sub-region of the very extensive W51 star-forming complex. \citet{2015PASJ...67...65N} derived a distance of $5.62^{+0.59}_{-0.49}$\,kpc to this region based upon parallax observations of water masers in G48.99-0.30 (which we adopt in this work). Even though G48.99-0.30 lies $\sim$3.5$\arcmin$ from the galactic coordinates of this region ($\ell$ = 48.930, $b$ = -0.286), G48.9-0.3 is quite extended in both the radio ($D_{6cm}=4.4\arcmin$) and mid-infrared ($D_{MSX22\mu m}$ $\sim$ 9$\arcmin$). Thus G48.99-0.30 is likely to be at the same distance as the rest of the G48.9-0.3 region. Indeed the list of v$_{lsr}$ values compiled from multiple studies by \citet{2014AA...569A.125H} for G48.99-0.30 (v$_{lsr}= 63-67$\,km/s) almost exactly matches those for G48.9-0.3 (v$_{lsr}= 64-67$\,km/s).

\textbf{W51A:} W51A contains two G\ion{H}{2} regions, G49.5-0.4 and G49.4-0.3, separated by $\sim$6$\arcmin$, both of which were covered in-depth in our \citetalias{2019ApJ...873...51L}. Trigonometric maser parallaxes were first measured towards G49.5-0.4 by \citet{2009ApJ...693..413X} using methanol masers, yielding a distance of $5.1^{+2.9}_{-1.4}$\,kpc. These were followed by measurements of the water maser parallaxes by \citet{2010ApJ...720.1055S} who obtained a much more precise values of $5.41^{+0.31}_{-0.28}$\,kpc. We adopt this value for the distance to W51A:G49.5-0.4. 

There have been no maser parallax measurements towards  W51A:G49.4-0.3, but we adopt the same distance as G49.5-0.4, since they both display similar v$_{lsr}$ values. \citet{2014AA...569A.125H} compiled a list of v$_{lsr}$ values from multiple studies and find a range of v$_{lsr}=56-59$\,km/s for G49.4-0.3, which is comparable to v$_{lsr}$$=53-55$\,km/s for G49.5-0.4 \citep[with an outlier of 67.38\,km/s in the H91$\alpha$ measurement of][]{2006ApJ...653.1226Q}.

\textbf{K3-50:} This region is referred to as W58A in \citet{2004MNRAS.355..899C}. The distance to this region has been kinematically derived by multiple studies, with all derived values falling in the range of 7.3 to 9.3\,kpc (e.g., \citealt{1975MNRAS.170..139H}; \citealt{2011ApJ...736..149G}; \citealt{2011ApJ...738...27B}; also see discussion in \citealt{2015MNRAS.453.2622B}), which places this region just outside the solar circle (with $R_{GC}>8.3$\,kpc) and in the outer galaxy. The v$_{lsr}$ measurements compiled by \citet{2014AA...569A.125H} show a range between $-23.11$ to $-26.22$\,km/s. The value from the H91$\alpha$ measurement of \citet{2006ApJ...653.1226Q} of v$_{lsr}=-23.11\pm0.08$\,km/s has the lowest error and yields a kinematic distance of $7.64^{+0.81}_{-0.54}$\,kpc, which we adopt here. One exception is from \citet{2011AA...532A.127D} who measure a line at a comparable velocity ($-23.3$\,km/s) but give a distance of 2.83\,kpc, assigning it to the kinematic tangent point, which seems unlikely. \citet{2010ApJ...714.1015S} were able to spectrally classify the star responsible for ionizing the K3-50D \ion{H}{2} region (as an O4V star) and based upon its brightness and estimated extinction a distance of $8.5^{+1.5}_{-0.6}$\,kpc was derived, which is inconsistent with the tangent point distance, and agrees with our more precise adopted distance to within the combined errors.

\textbf{DR7:} Like K3-50, most previous observations of DR7 have yielded kinematic distances around $7-8$\,kpc, which places it just outside the solar circle and in the outer galaxy. \citet{2014AA...569A.125H} compiled a list of v$_{lsr}$ values from multiple studies and find a range of $-42<v_{lsr}<-37$\,km/s. Measurement in the H91$\alpha$ transition by \citet{2006ApJ...653.1226Q} yield a v$_{lsr}=-39.17\pm$0.07, which gives a kinematic distance of $7.30^{+0.84}_{-0.72}$\,kpc. \citet{2011AA...532A.127D} also measure a line at $-41$\,km/s, but assign the source to the tangent point distance of 1.56\,kpc which seems unlikely. With no spectrophotometric or maser parallax observations available, we adopt the kinematic distance of 7.30\,kpc.

\textbf{W3:} W3 is located in the nearby Perseus Arm. Sources within this arm tend to have peculiar velocities, as evidenced by the difference in distance when derived via kinematic measurements \citep[$\sim$4\,kpc, like that adopted by][]{2004MNRAS.355..899C} compared to $\sim$2\,kpc from spectrophotometric and trigonometric parallaxes \citep{2019MNRAS.487.2771N}. A trigonometric maser parallax was measured towards W3(OH) by \citet{2006Sci...311...54X} who derive a distance of 1.95$\pm$0.04\,kpc, however this source is more than 16$\arcmin$ from W3 Main (which is the regions we are considering here). \citet{2019MNRAS.487.2771N} use GAIA parallax measurements to show that different parts of the W3 star-forming complex appear to have slighty different distances. They obtain a distance to W3(OH) of $2.00^{+0.29}_{-0.23}$\,kpc, which is consistent to within the errors with the maser-derived distances of \citet{2006Sci...311...54X}. However, for W3 Main \citet{2019MNRAS.487.2771N} obtain a distance of $2.30^{+0.19}_{-0.16}$\,kpc, which we will adopt here.

\textbf{RCW\,42:} This region is not well studied, and thus only kinematic distances are available. Using the H109+110$\alpha$ transition observations from \citet{1987AA...171..261C} of v$_{lsr}$=39.0$\pm$1.0\,km/s (the only observation from \citealt{2014AA...569A.125H} with reported errors), we derive a kinematic distance of $5.97^{+0.90}_{-0.72}$\,kpc and place it in the outer galaxy. This value is consistent to within the errors of the value of 6.4\,kpc adopted by \citet{2004MNRAS.355..899C}.

\textbf{RCW\,46:} This source is also known as IRAS 10049-5657. Kinematically-derived distance measurements vary between $\sim$5$-7$\,kpc \citep{2008AJ....136.1427V}, and this is due to the relatively large range in v$_{lsr}$ measurements toward this location \citep[$19.0-26.2$\,km/s;][]{2014AA...569A.125H}. The line measurement with the smallest error is 19.0$\pm$1.0\,km/s which yields a distance of $5.77^{+0.77}_{-0.77}$\,kpc, based upon the radio recombination line measurements of \citet{1987AA...171..261C}. This source lies in the outer galaxy, and therefore does not have a kinematic distance ambiguity. There does exist a spectrophotometric measurement of the distance towards this source by \citet{2011MNRAS.411..705M} who derive a value of 6.97$\pm$2.72\,kpc, however the errors are quite large. We therefore adopt the 5.77\,kpc value derived kinematically.

\textbf{RCW\,49:} As one of the most luminous G\ion{H}{2} regions in the southern hemisphere, RCW\,49 has been heavily studied. The stellar cluster Westerlund\,2 is believed to be contained within the \ion{H}{2} region of RCW 49 and responsible for its ionization. The recent work of \citet{2021ApJ...914..117T} discusses the considerable variation of the accepted distance to this region over the decades, and we refer the reader to that work for the details. We will follow the recommendation of \citet{2021ApJ...914..117T} and adopt in this work the spectrophotometrically-derived distance to Westerlund\,2 of 4.16$\pm$0.27\,kpc from \citet{2013AJ....145..125V} for RCW\,49.

\textbf{NGC\,3372:} This region is also known as the Carina Nebula. The luminous blue variable, $\eta$\,Carinae, is thought to be located within this nebula \citep{2008hsf2.book..138S}. From the expansion parallax of the Homunculus nebula around $\eta$\,Car an accurate distance of 2.3$\pm$0.1\,kpc was found to the star \citep{2006MNRAS.367..763S}, and we adopt that distance here.

\textbf{G289.066-0.357:} \citet{2009ApJ...699..469C} detected emission in the H166$\alpha$ transition from this region and determined a kinematic distance of 7.1$\pm$0.3\,kpc. This distance is consistent with the value we derive from the H109+110$\alpha$ transition measurements of \citet{1987AA...171..261C}, who find a v$_{lsr}$=19.0$\pm$1.0\,km/s, which we calculate to be $7.15^{+0.54}_{-0.93}$\,kpc (which we adopt here). This source is outside the solar circle, and therefore there is no near/far distance ambiguity. One outlier measurement is the distance of 3.1\,kpc derived by \citet{2018MNRAS.476..842O} using a color-magnitude diagram analysis of the 2MASS sources found in the vicinity of G289.066-0.357. Such a near distance may mean the NIR stars used in the analysis are not associated with the radio region at all but instead are in the foreground.

\textbf{NGC\,3576:} This source is also known as RCW 57 and IRAS 11097-6102. \citet{2018ApJ...864..136B} have determined GAIA parallax observations towards stars believed to be associated with the ionized radio emission of NGC\,3576. The value of 2.77$\pm$0.31\,kpc is close to the previously derived kinematic distances \citep[e.g.,][]{1999AJ....117.2902D} which place it to be at or near the tangent point of $\sim$3.0\,kpc.  

Contrarily, \citet{2011MNRAS.411..705M} spectrophotometrically derive a distance of 0.98$\pm$0.19\,kpc. Given that the distance derived via GAIA parallaxes is thought to be a more accurate method and it is consistent with the kinematic distance, we adopt that value in this work.

\textbf{NGC\,3603:} There is a relatively large range of kinematic distances toward this region found in the literature ranging between $6-10$\,kpc \citep{2019AA...625L...2K}. Consistent with this, \citet{1999AJ....117.2902D} found that the H90$\alpha$ velocities measured toward thirteen sub-regions within NGC\,3603 range between $-2$\,km/s and +19\,km/s. They estimate a kinematic distance of 6.1$\pm$0.6 kpc based upon the line velocity of v$_{lsr}$ = 9.1\,km/s found by integrating over the entire range of line velocities. \citet{2008AJ....135..878M} performed spectroscopic parallax measurements towards multiple massive stars within NGC\,3603 to derive a distance of 7.6\,kpc. More recently, \citet{2019MNRAS.486.1034D} used GAIA parallaxes towards cluster members to derive a distance of 7.2$\pm$0.1\,kpc which we adopt here.

\textbf{G298.227-0.340/G298.862-0.438:} These two sources ($\sim$0.6$\arcdeg$ apart) are the two brightest \ion{H}{2} regions within a much larger star-forming complex called the Dragonfish Nebula \citep{1997AA...319..788R} that resides in the outer Galaxy at a kinematically derived distance of $\sim$10\,kpc. For example, using the v$_{lsr}$ measured by \citet{1987AA...171..261C} of 31.0$\pm$1.0\,km/s for G298.227-0.340 and 25.0$\pm$1.0\,km/s for G298.862-0.438 we obtain distances of $10.40^{+0.66}_{-0.66}$\,kpc and $10.02^{+0.65}_{-0.58}$\,kpc, respectively. This similarity indicates the two regions are likely part of the same physical region. \citet{2016AA...589A..69D} argue that the stellar cluster Mercer 30 is related to the Dragonfish Nebula and use spectrophotometric techniques to determine its distance as being 12.4$\pm$1.7\,kpc, which is consistent with the kinematic distances to within the errors. We adopt that distance here for both sources. Contrarily, \citet{2011MNRAS.411..705M} derive a spectrophotometric distance of 4.73$\pm$1.78\,kpc towards G298.227-0.340, which seems inconsistent with all other measurements.

\textbf{G305.359+0.194:} \citet{2018ApJ...864..136B} have determined a distance to this region using GAIA parallax observations towards stars believed to be associated with the G305.359+0.194 star-forming region. Their value of 3.59$\pm$0.85\,kpc agrees to within the errors with previously derived kinematic near distances of $\sim$3.4\,kpc \citep[e.g.,][]{2015ApJ...806..199B}. 

\textbf{G319.158-0.398:} Values of the v$_{lsr}$ toward this region range between $-16$ and $-27$\,km/s \citep[e.g.,][]{2018MNRAS.473.1059U, 2014AA...569A.125H} which yield near/far kinematic values of $\sim$1.5/11.0\,kpc. \citet{2012MNRAS.420.1656U} measure \ion{H}{1} absorption toward this region and determine that the region is likely at the far kinematic distance. Using the measurement with the lowest error of $-21.0\pm$1.0\,km/s from the H109+110$\alpha$ observations of \citet{1987AA...171..261C} yields a far kinematic distance of $11.26^{+0.35}_{-0.42}$\,kpc, which we adopt here.

\textbf{G319.392-0.009:} Values of the v$_{lsr}$ toward this region range between $-11$ and $-19$\,km/s \citep[e.g.,][]{2018MNRAS.473.1059U, 2014AA...569A.125H}, yielding near/far kinematic values of $\sim$1.0/11.5\,kpc. Using the measurement with the lowest error of $-14.0\pm$1.0\,km/s from the H109+110$\alpha$ observations of \citet{1987AA...171..261C} yields near/far kinematic distances of
$1.01^{+0.23}_{-0.43}$/$11.78^{+0.34}_{-0.42}$\,kpc. Since \citet{2012MNRAS.420.1656U} determine this region is at its far kinematic distance due to their measurements of \ion{H}{1} absorption, we adopt the far kinematic distance of 11.78\,kpc. 

\textbf{G320.327-0.184:} This source is also known as IRAS\,15061-5806. \citet{2018MNRAS.473.1059U} catalog five ATLASGAL clumps within 2$\arcmin$ of the galactic coordinates for this region with v$_{lsr}$ values ranging between $-6.9$ and $-11.1$\,km/s. They report that the kinematic distance ambiguity toward this region is resolved via \ion{H}{1} absorption observations, which point to the near distance. This is contrary to the far distance listed by \citet{2004MNRAS.355..899C} which is reported from \citet{2003AA...397..133R} who observe CO features up to $-69$\,km/s. We adopt the near distance here of $0.64^{+0.38}_{-0.27}$\,kpc based upon the v$_{lsr}$ measurement of $-11.0\pm$1.0\,kpc from \citet{1987AA...171..261C}, however we treat this source distance as still ambiguous given the conflicting absorption measurements.

\textbf{RCW\,97:} The large G\ion{H}{2} region of RCW\,97 lies in the northern part of the G327.293-0.579 molecular cloud which also contains an infrared dark cloud to the south \citep{2006AA...454L..91W}. Only kinematic estimates are available for the distance to this source, with line velocities in many transitions having a range of $-43.0 > v_{lsr} > - 49.8$\,km/s \citep[e.g.,][]{2021ApJS..253....2H,2014ApJS..212....2G,2006AA...454L..91W,2014AA...569A.125H}. Both \citet{2014ApJS..212....2G} and \citet{2012MNRAS.420.1656U} determine this region is at its near kinematic distance due to the presence of absorption features like \ion{H}{1}. Using the v$_{lsr}$ measurement of $-47.5\pm$0.1\,kpc from C$^{18}$O measurements of \citet{2006AA...454L..91W} yields our adopted distance of $2.98^{+0.23}_{-0.36}$\,kpc.

\textbf{G327.993-0.100:} \citet{2018MNRAS.473.1059U} find three ATLASGAL clumps here within 45$\arcsec$ of each other and all three are contained in the infrared emitting area as seen in the MSX data from \citet{2004MNRAS.355..899C}. \citet{2018MNRAS.473.1059U} claim from \ion{H}{1} absorption analyses that two of these clumps have the larger kinematic distance of $\sim$11\,kpc, while one has the near kinematic distance of $\sim$3\,kpc. However, they also claim that these sources are all part of a cluster which is at 3.1\,kpc. Further evidence that the region is at the near distance comes from \citet{2004MNRAS.347..237P}, who claim this region is at the near kinematic distance because it has an optical counterpart. We will therefore adopt the near distance in this work as well. Using the v$_{lsr}$ measurement of $-45\pm1.04$ from the H109+110$\alpha$ of \citet{1987AA...171..261C} yields a distance of $2.80^{+0.31}_{-0.31}$\,kpc. 

\textbf{G330.868-0.365:} The range of v$_{lsr}$ measurements to this region lie between $-56.0$ and $-63.3$\,km/s. \citet{2012ApJ...753...62J}, \citet{2004MNRAS.347..237P}, and \citet{2012MNRAS.420.1656U} all determine a nearby distance from \ion{H}{1} absorption observations. Using the value of $-56.0\pm1.0$\,km/s from the H109+110$\alpha$ observations of \citet{1987AA...171..261C} yields a near distance of $3.44^{+0.37}_{-0.30}$\,kpc, which we adopt here. Interestingly, \citet{2004MNRAS.355..899C} use the far distance claimed by \citet{2003AA...397..133R}, but it is unclear why the far distance is adopted when all of their measured absorption features all have velocities similar to the v$_{lsr}$ range mentioned above.

\textbf{G331.324-0.348:} \citet{2018MNRAS.473.1059U} shows four ATLASGAL sources all with similar velocity ($-66.4 < v_{lsr} < -65.5$\,km/s) all claimed to be in the same cluster at the near kinematic distance due to the presence of \ion{H}{1} self absorption. \citet{2004MNRAS.347..237P} also claim the near distance from their \ion{H}{1} observations. \citet{2004MNRAS.355..899C} use the far distance adopted by \citet{2003AA...397..133R}, because they claim to see CO absorption at a very different velocity ($-99$\,km/s). Spectrophotometric observations by \citet{2012MNRAS.423.2425P} derive a distance of 3.29$\pm$0.58\,kpc, consistent with the near kinematic distance. We adopt the distance of \citet{2012MNRAS.423.2425P} in this work.

\textbf{G331.354+1.072:} The H109+110$\alpha$ observations of \citet{1987AA...171..261C} toward this region yield a v$_{lsr}$ of $-79\pm1.0$\,kpc, consistent with other measured transitions. \citet{2018MNRAS.473.1059U} find one ATLASGAL clump with 2$\arcmin$ of the galactic coordinates of this region and measure a v$_{lsr}$ of $-78.3$\,km/s and assume near kinematic distance due to this region's high galactic latitude. Though the \ion{H}{1} absorption observations from \citet{2012ApJ...753...62J} were ambiguous, \citet{2012MNRAS.420.1656U} measure \ion{H}{1} absorption consistent with the near distance as well. We therefore adopt the near distance in this work, and using the v$_{lsr}$ of \citet{1987AA...171..261C}, derive a distance of $4.50^{+0.55}_{-0.34}$\,kpc.

\textbf{G331.529-0.084:} Only kinematic measurements have been made towards this region and the distance is highly uncertain. \citet{2018MNRAS.473.1059U} find four ATLASGAL molecular clumps within 2$\arcmin$ of the radio coordinates for this source, all with comparable v$_{lsr}$ values ranging from $-87.3$ to $-89.2$\,km/s. However, only one of these clumps has \ion{H}{1} absorption measurements consistent with the far distance, while the remaining three are claimed to be too ambiguous to solve. Nonetheless, \citet{2018MNRAS.473.1059U} considers the source as part of the same cluster and places them at the near kinematic distance. \citet{2014ApJS..212....2G} state they resolved the distance ambiguity to be near based upon H$_2$CO absorption measurements, but \citet{2004MNRAS.347..237P} claim the far kinematic distance based upon \ion{H}{1} absorption measurements. Yet others \citep[e.g.,][]{2012ApJ...753...62J} claim that the source is at the kinematic tangent point. \citet{2013ApJ...774...38M} describe the prior distance measurements in detail and decide that the region is likely at the tangent point as well and choose to use a large uncertainty of 30\% given by the near and far kinematic distances. We will employ that strategy here. The v$_{lsr}$ measurements compiled by \citet{2014AA...569A.125H} show a range between $-88.4$ and $-90.6$\,km/s, with an outlier of $-100.7$\,km/s from the CS observations of \citet{1996AAS..115...81B}. Using the H109+110$\alpha$ observations of \citet{1987AA...171..261C} of v$_{lsr}$ of $-89\pm1.0$\,kpc, gives a near distance of $5.11^{+0.60}_{-0.47}$\,kpc and a far distance $9.48^{+0.50}_{-0.58}$\,kpc. However, we will adopt the tangent point distance of 7.31$\pm$2.19\,kpc, with the large errors to reflect the near/far distances. Despite all of this doubt in distance, we will point out that that this region is so bright that even at the near kinematic distance, G331.529-0.084 would still qualify as a G\ion{H}{2} region. 

\textbf{G333.122-0.446:} This is the first of three sources on our list that are believed to be co-located within the G333 giant molecular cloud, one of the most massive in the fourth quadrant of the Galaxy \citep{2016MNRAS.458.3429W}. \citet{2018MNRAS.473.1059U} find three ATLASGAL sources, assumed to be in a cluster at near kinematic distance. The \ion{H}{1} absorption measurements of \citet{2012ApJ...753...62J} and \citet{2012MNRAS.420.1656U} are consistent with a near kinematic distance, and \citet{2004MNRAS.347..237P} claim the near distance because this \ion{H}{2} region has an optical component. Given the v$_{lsr}$ of this region \citep[e.g., $-52\pm1.0$\,km/s from][]{1987AA...171..261C} a near kinematic distance would be $\sim$3.3\,kpc. However, an even nearer and more accurate distance was measured by \citet{2005AJ....129.1523F} via spectrophotometric techniques of 2.6$\pm$0.2\,kpc. Later spectrophotometric measurements by \citet{2011MNRAS.411..705M} yielded a value of 3.57$\pm$1.33\,kpc, which are consistent with the value from \citet{2005AJ....129.1523F} to within the errors. As even \citet{2011MNRAS.411..705M} points out, the results from \citet{2005AJ....129.1523F} are more reliable, and thus we adopt their distance here. This distance is also consistent with the distances measured toward the other two sources in the G333 complex (G333.293-0.382 and G333.610-0.217 discussed below), signifying that they may indeed be related.

\textbf{G333.293-0.382:} This is the second of three sources on our list that are believed to be reside within the G333 giant molecular cloud. \citet{2018MNRAS.473.1059U} find three ATLASGAL source within 2$\arcmin$ of the galactic coordinates of this source, and claim they are in a cluster at the near kinematic distance due to the \ion{H}{1} absorption observations of \citet{2012MNRAS.420.1656U}. \citet{2014ApJS..212....2G} and \citet{2004MNRAS.347..237P} also both claim the near distance as well due to the presence of an optical component associated with the \ion{H}{2} region here. We derive a near distance of $\sim$3.2\,kpc towards this region using typical v$_{lsr}$ values toward this region \citep[e.g., $-50\pm1.0$\,km/s from][]{1987AA...171..261C}. At this distance G333.293-0.382 just makes it over the criterion of being a G\ion{H}{2} region (log$N_{LyC}=50.00$\,photons/s). However, the spectrophotometric results of \citet{2009MNRAS.394..467R} towards the associated source IRAS 16177–5018–IRS1 derive a distance of at most 2.6$\pm$0.7\,kpc given the best match of their spectra to a O3If\text{*} supergiant. However, they point out that such stars in star-formation regions are rare, and that the distance would be 1.2$\pm$0.7\,kpc if one assumes the source to instead be a O3V-O5V main sequence star. This closer distance is much smaller than the kinematic distance, however either of these two distances would put G333.293-0.382 out of contention for being a G\ion{H}{2} region. We will adopt here the larger spectrophotometric distance (being that it is closer to the kinematic distance) of 2.6\,kpc, which leads to a log$N_{LyC}=49.79$\,photons/s and is consistent with the distances measured to the two other regions within the G333 complex, G333.122-0.446 and G333.610-0.217.

\textbf{G333.610-0.217:} This is the third of three sources on our list that are believed to be co-located within the G333 giant molecular cloud, and is the most prominent and best studied among the three \citep[e.g.,][]{2020AA...633A.155R, 2014ApJS..213....1T}. This source is associated with the \ion{H}{2} region known as RCW\,106. \ion{H}{1} absorption measurements of \citet{2004MNRAS.347..237P}, and \citet{2012MNRAS.420.1656U} point to a near kinematic distance. However, \citet{2020AA...633A.155R} derive a distance of 2.54$\pm$0.71\,kpc from GAIA parallax measurements of cluster members within G333.6-0.2, which we adopt here. This distance is in agreement with the reported distances of both G333.122-0.446 and G333.293-0.382 above, again implying they are all indeed co-located within the G333 giant molecular cloud. 

\textbf{G338.398+0.164:} The galactic coordinates of this source are close to that of the star cluster Mercer 81 ($\ell$=338.384, $b$=0.111) towards which \citet{2012MNRAS.419.1860D} performed a kinematic analysis placing the cluster close to where the far end of the Galactic Bar intersects the Norma spiral arm at a distance of 11$\pm$2\,kpc. There is still some confusion regarding distance, however shown by \citet{2018MNRAS.473.1059U} who find four ATLASGAL source within 2$\arcmin$ of the galactic coordinates of this source, and claim they are in a cluster at the near kinematic distance (2.7\,kpc) due to the \ion{H}{1} absorption observations. However of the four ATLASGAL sources, the one closest to the galactic coordinates of G338.398+0.164 is shown by \citet{2012MNRAS.420.1656U} to have \ion{H}{1} absorption in keeping with the far distance, one source is thought to be at the near distance, and the remaining two are too ambiguous to decide. Given all of the uncertainty, we will tentatively adopt the far distance indicated by the ATLASGAL source closest to the coordinates of this source. Using the value of $-29.0\pm$1.0\,km/s from the H109+110$\alpha$ observations of \citet{1987AA...171..261C} we derive a far distance of $13.29^{+0.25}_{-0.45}$\,kpc, which we adopt here.

\textbf{G338.400-0.201:} This region is not studied very well, and there are only a handful of observations from which to assess the distance (with conflicting results). \citet{1970AA.....6..364W} find a value of v$_{lsr}$ = $-4.3\pm3.6$\,km/s in the H109$\alpha$ transition, while \citet{1987AA...171..261C} measure 2.0$\pm$1.0\,km/s in the similar H109+110$\alpha$ transition. Meanwhile \citet{2007AA...474..891U} measure a $^{13}$CO v$_{lsr}$ of 4.0\,km/s. At positive velocities, the region exists in the outer galaxy \citep[as argued by][]{2012ApJ...753...62J}, and thus has no distance ambiguity. For v$_{lsr}$ values at the more negative end of the measured range, there is a chance the source could have a very close ($<0.5$\,kpc) near kinematic distance. The MSX 22$\mu$m images of \citet{2004MNRAS.355..899C} show a relatively faint source at this location, with a FWHM$<1\arcmin$. Such an underwhelming infrared region may be indicating that the source is a modest \ion{H}{2} region at the near distance and not a distant G\ion{H}{2} region, which tend to have more complex and extended morphologies (as discussed for Sgr\,D and W42 in Section~\ref{sec:reject}). However, without more evidence we will not dismiss the source as a candidate G\ion{H}{2} region and will tentatively use the \citet{1987AA...171..261C} v$_{lsr}$, which leads to an outer galaxy kinematic distance of $15.71^{+0.58}_{-0.40}$\,kpc.

\textbf{G345.555-0.043:} This region is known as the G345.555-0.042 GMC complex \citep{2012MNRAS.420.1656U}, but is not well studied. There are two ATLASGAL sub-millimeter clumps within 2$\arcmin$ of these galactic coordinates of this region, both of these clumps lie $\sim$1.7$\arcmin$ away. \citet{2003AA...397..133R} and \citet{1987AA...171..261C} find CO and H$_2$CO transitions at much higher absolute velocities than the v$_{lsr}$= $-6.0\pm1.0$\,km/s derived from H109+110$\alpha$ observations of \citet{1987AA...171..261C}, and thus claim the far kinematic distance to this region. Using that v$_{lsr}$ value we derive a far distance of $15.28^{+0.57}_{-0.35}$\,kpc, which we adopt here. 

\textbf{G345.645+0.009:} This source has a radio-emitting size of $\theta_{6cm} = 4.2\arcmin$ according to \citet{2004MNRAS.355..899C}. \citet{2018MNRAS.473.1059U} find four ATLASGAL molecular clumps within this radio source area, all with comparable v$_{lsr}$ values ($-6.0$ to $-10$\,km/s). However, two of these clumps have \ion{H}{1} absorption measurements consistent with the near distance and two have \ion{H}{1} absorption measurements consistent with the far distance. The infrared source at the center of the radio emission is only 1$\farcm$9 in diameter in the 22$\mu$m MSX images \citep{2004MNRAS.355..899C}, and the only one ATLASGAL clump is contained within that region, and that clump is one determined to be at the far kinematic distance due to \ion{H}{1} absorption measurement \citep{2012MNRAS.420.1656U}. Using the v$_{lsr}$= $-10.0\pm1.0$\,km/s value derived from H109+110$\alpha$ observations of \citet{1987AA...171..261C} we derive a far distance of $14.97^{+0.39}_{-0.45}$\,kpc, which we adopt here.

\textbf{G347.611+0.204:} This source has a 6\,cm radio-emitting diameter of $6.1\arcmin$ according to \citet{2004MNRAS.355..899C}. Like our previous example, \citet{2018MNRAS.473.1059U} find multiple ATLASGAL molecular clumps within this radio source area, totalling five clumps all with comparable v$_{lsr}$ values ($-91$ to $-97$\,km/s). One of these clumps has \ion{H}{1} absorption measurements consistent with the near distance, one with the far distance, and three are too ambiguous to determine. The stellar cluster [DBS2003]\,179 \citep{2003AA...400..533D} is believed to be associated with the \ion{H}{2} emission here. \citet{2012AA...546A.110B} provides a good overview of the history of distance measurements towards G347.611+0.204 and derive a spectrophotometric distance of 7.9$\pm$0.8\,kpc, which we adopt here. This distance is consistent with the near kinematic distance of $\sim$7.0\,kpc.

\textbf{G351.467-0.462:} \citet{2018MNRAS.473.1059U} find 3 ATLASGAL sources within 2$\arcmin$ of these galactic coordinates all at same velocity (v$_{lsr}$ = $-22$ to $-23$\,km/s) and assume these to be a related cluster at a near kinematic distance due to \ion{H}{1} absorption. This is supported by similar measurement by \citet{2006ApJ...653.1226Q}, though \citet{2013ApJ...774..117J} say their \ion{H}{1} absorption spectrum does no conclusively point one way or the other. \citet{2006AA...455..923B} derive a spectrophotometric distance to the region by studying a stellar cluster that is coincident with the ionized gas in this region. They calculate a distance $\sim$3.2\,kpc but report no formal no uncertainty. Using the v$_{lsr}$ measurements of the H91$\alpha$ transition towards this source by \citet{2006ApJ...653.1226Q} of $-21.44\pm0.74$\,km/s yields a near kinematic distance of $3.24^{+0.34}_{-0.26}$\,kpc, which we will adopt here because it is consistent with the spectrophotometric distance determination.

\textbf{Sgr\,C:} As pointed out by \citet{2013ApJ...775L..50K}, the v$_{lsr}$ of Sgr\,C is very similar to those of sources in the Near 3 kpc Arm at a distance of $\sim$5.5 kpc, which complicates kinematic interpretations. That being said, measured v$_{lsr}$ values, like that of \citet{1987AA...171..261C} ($-60\pm1.0$\,km/s from the measured H109+110$\alpha$ transitions) yield tangent point kinematic distances ($8.34^{+0.15}_{-0.17}$\,kpc) that agree with the distance to the Galactic Center to within the errors.

\end{document}